\documentclass[amsmath,amssymb,reprint,groupedaddress]{revtex4-1}

 \pdfoutput=1    

\usepackage{graphicx}
\usepackage{dcolumn}
\usepackage{bm}
\usepackage{siunitx}
\usepackage{lipsum}
\usepackage{enumitem}
\usepackage{relsize}
\usepackage{threeparttable}
\usepackage{multirow}
\usepackage{tikz}

\newcommand{\vek}[1]{\boldsymbol{#1}}    
\newcommand{\ffrac}[2]{\frac{\displaystyle #1}{\displaystyle #2}}
\newcommand{\dif}{\mathrm{d}}       
\newcommand{\me}{\mathrm{e}}    
\newcommand{\eexp}[1]{\me^{\displaystyle #1}}

\begin{document}
\title{
Identifying the Onset of Phase Separation in Quaternary Lipid Bilayer Systems \\ from Coarse-Grained Simulations
}
\author{Shushan He}
\author{Lutz Maibaum}
\affiliation{Department of Chemistry, University of Washington, Seattle, WA 98195 }
\begin{abstract}

Understanding the (de)mixing behavior of multicomponent lipid bilayers is an important step towards unraveling the nature of spatial composition heterogeneities in cellular membranes and their role in biological function. We use coarse-grained molecular dynamics simulations to study the composition phase diagram of a quaternary mixture of phospholipids and cholesterol. This mixture is known to exhibit both uniform and coexisting phases. We compare and combine different statistical measures of membrane structure to identify the onset of phase coexistence in composition space. An important element in our approach is the dependence of composition heterogeneities on the size of the system. While homogeneous phases can be structured and display long correlation lengths, the hallmark behavior of phase coexistence is the scaling of the apparent correlation length with system size. Because the latter cannot be easily varied in simulations, our method instead uses information obtained from observation windows of different sizes to accurately distinguish phase coexistence from structured homogeneous phases. This approach is built on very general physical principles, and will be beneficial to future studies of the phase behavior of multicomponent lipid bilayers.

\end{abstract}

\pacs{Valid PACS appear here}
\maketitle

\section{\label{sec:intro}Introduction}

The role of spatial composition heterogeneities in biological membranes is a long-standing problem in membrane biophysics~\cite{Simons1997,Simons2004,Simons2011,Lingwood2010,Anderson2002,Toulmay2012,Diaz-Rohrer2014a,Levental2016}. The physical mechanism that gives rise to these heterogeneities as well as their role in biological function remain poorly understood~\cite{Rajendran2005,Simons2010,Diaz-Rohrer2014a}. Phospholipids, as major constituents of biological membranes, are believed to contribute to composition heterogeneity through their complex phase behavior~\cite{Veatch2005,Ingolfsson2014}. It is well known  that under some conditions model membrane systems can exhibit spatial heterogeneity at the micrometer length scale~\cite{Veatch2002,Veatch2003,Veatch2005,Zhao2007,Konyakhina2011,Toulmay2013,Diaz-Rohrer2014}. This micron-scale heterogeneity is characterized by the formation of two distinct lipid phases both in perturbed cell membranes and in model membranes~\cite{Holowka2005,Sengupta2007}: liquid-ordered ($L_o$) regions are enriched in cholesterol and lipids with saturated tails and high melting temperatures, while lipids with unsaturated tails and lower melting points are typically found in liquid-disordered ($L_d$) regions. These regions correspond to two coexisting thermodynamic phases, as shown for example by fluorescence microscopy experiments on giant unilamellar vesicles (GUVs)~\cite{Veatch2003}: starting from a homogeneous state at high temperature, the membrane spontaneously separates into liquid-ordered and liquid-disordered domains when cooled below a characteristic transition temperature, and these domains then diffuse and coalesce until complete phase separation is achieved. A multitude of experiments employing different techniques and membrane compositions have established that many model membranes can undergo phase separation into coexisting $L_o$ and $L_d$ regions~\cite{Baumgart2003, Veatch2005,Zhao2007,Marsh2009,Heberle2010}.

In cellular plasma membranes no such large domains have been observed with conventional microscopy. However, experiments on cell-derived Giant Plasma Membrane Vesicles (GPMVs) have shown the existence of liquid-order-like and liquid-disorder-like domains \cite{Baumgart2007, Veatch2008} and that such domains contribute to cell functions such as protein sorting \cite{Diaz-Rohrer2014}. With experimental techniques such as F\"orster resonance energy transfer (FRET), neutron or X-ray scattering, and super-resolution fluorescence, researchers have detected characteristic signals that are consistent with the existence of heterogeneity on nanometer length scales on unperturbed membranes of live cells \cite{Sengupta2007,Sengupta2007a,Edidin2003, Stone2017}. 

To reconcile the discrepancy in length scale of lipid spatial heterogeneity between different membrane systems and the connection between membrane heterogeneity and lipid composition, numerous studies have been performed on model membrane systems, especially unilamellar vesicles, with well-controlled lipid compositions. These experiments have provided insight into the mechanism that controls the length scale of heterogenous domain formation using fluorescence microscopy, FRET, NMR, or other techniques \cite{Veatch2003, Heberle2010, Konyakhina2011, Petruzielo2013,Konyakhina2013, Heberle2013a, Lindblom2009, Leftin2014, Grelard2010, Ando2015}.  It has been observed that model membrane systems composed of ternary mixtures of dipalmitoylphosphatidylcholine (DPPC), 1,2-dioleoyl-sn-glycero-3-phosphocholine (DOPC), and cholesterol show micron-scale heterogeneity at optical resolution, characterized as $L_d$--$L_o$ phase coexistence below the miscibility temperature \cite{Veatch2003}. However, a mixture of DPPC, cholesterol and 1-palmitoyl-2-oleoyl-sn-glycero-3-phosphocholine (POPC) did not exhibit optically observable lipid heterogeneity, while FRET and electron spin resonance (ESR) experiments suggest that heterogenities exist on nanometer length scales~\cite{Heberle2010}.

To further probe this difference in length scale of membrane heterogeneity, four-component lipid mixtures were studied \citep{Goh2013, Konyakhina2013, Ackerman2015}. Here the onset of lipid heterogeneity can be investigated by manipulating the compositions of two unsaturated lipids (POPC and DOPC). Microscopy experiments on GUVs reveal the existence of an additional phase, characterized by stripe-like features on the vesicle surface, inbetween the phase-separated and the homogeneous (but nanoscopically ordered) regions~\cite{Konyakhina2011}. Several models have been proposed to rationalize the emergence of such stripe-like morphologies, including the competition between line tension and curvature effects~\cite{Goh2013}, critical fluctuations~\cite{Veatch2008, Honerkamp-Smith2009}, and the existence of a nearby microemulsion phase~\cite{Shlomovitz2013}. Interestingly, these features were not observed in neutron scattering experiments on much smaller vesicles~\cite{Heberle2013}.

Due to the elusive nature of sub-micron scale domains and difficulties in experimentally resolving and interpreting scattering data, computer simulations of multicomponent bilayers provide a promising approach to study their spatial organization. Along with previous simulations of cholesterol-containing membranes \cite{Niemela2007,Rog2009,Perlmutter2009,Pandit2004}, an all-atom membrane simulation on the microsecond time scale suggests that cholesterol plays an important role in $L_o$--$L_d$ phase coexistence in a ternary lipid mixture~\cite{Sodt2015}. In addition, several groups have used the coarse-grained MARTINI model \cite{Marrink2004, Marrink2007} to simulate such membranes, and have confirmed that this model can qualitatively reproduce experimentally observed phase behavior~\cite{Risselada2008,Rosetti2012, Baoukina2013,Ackerman2015}. It is worth noting that many of these studies use the polyunsaturated lipid ,2-dilinoleoyl-sn-glycero-3-phosphocholine (DUPC or DLiPC) instead of DOPC because the MARTINI model fails to reproduce the experimentally observed phase separation of DPPC/DOPC/cholesterol membranes~\cite{Davis2013}.

In this work we present results from coarse-grained Molecular Dynamics (MD) simulations. We focus on a quaternary mixture of phosphatidylcholine (PC) lipids and cholesterol with different levels of unsaturation. Specifically, we use fully saturated DPPC, one-chain-singly-unsaturated POPC, doubly-unsaturated DUPC, and cholesterol (CHOL) in a planar model lipid bilayer. We employ a statistical mechanics approach to investigate the phase coexistence behavior and the structural properties of nanoscopic and intermediate regimes of this mixture.  We find a transition between the homogeneous and phase-separated regimes as we increase the DUPC content of the system. For the latter we show how one can obtain the compositions of the coexisting liquid-ordered and liquid-disordered phases both from partial density correlation functions and from local density distribution functions. To characterize the nature and length scale of spatial heterogeneity for membrane systems and to identify the onset of lipid phase segregation, we apply these approaches to a series of quaternary lipid mixtures at various unsaturated lipid content, as well as to systems of different sizes. We observe that the difference in composition between lipid domains is proportional to the global composition of the doubly unsaturated lipid (DUPC). Furthermore, we show that local density and composition distribution functions strongly depend on the size of the observation window chosen for the analysis, which allows us to quantify the length scale of heterogeneity. This length scale then plays a crucial role in identifying the nature of the heterogeneity. The hallmark feature of a system undergoing phase coexistence is that the length scale of heterogeneity is proportional to the size of the system, while a homogeneous system does not show such a dependence even in the presence of long-ranged correlations. Finally, we combine the results from these different analysis methods to identify the onset of phase separation in this quaternary lipid mixture.

\section{\label{sec:methods}Methods}
\subsection{Molecular Dynamics Simulation}
\subsubsection{Force Field}
We use the coarse-grained MARTINI 2.0 force field, which speeds up bilayer simulations by as much as 3-4 orders of magnitude compared to atomistic models \cite{Marrink2004,Marrink2007}. 
Explicit solvent representation was included in the force field with a four-to-one mapping of water molecules into a MARTINI water bead. 

\subsubsection{Bilayer Composition}

Our choice of quaternary mixture is inspired by the experimental and computational studies of Refs.~\onlinecite{Konyakhina2011} and \onlinecite{Ackerman2015}, respectively. We study symmetric lipid bilayers that consist of one third DPPC molecules, one third cholesterol molecules, and the remaining third is made of varying amounts of POPC and DUPC molecules. To vary the lipid composition and observe its effect on membrane phase behavior we move along the POPC/DUPC binary axis with the compositions of the other two species fixed. We define the composition variable
\begin{equation}
	\chi =  \ffrac{\left[\mathrm{DUPC}\right]}{\left[\mathrm{DUPC}\right] + \left[\mathrm{POPC}\right]}
\end{equation}
where $\left[\mathrm{DUPC}\right]$, $\left[\mathrm{POPC}\right]$ denote the global partial lipid densities. 
We vary $\chi$ by changing the relative composition of POPC and DUPC while keeping $\left[ \mathrm{DPPC} \right]$ and $\left[ \mathrm{CHOL} \right]$ constant. 

By changing $\chi$ from 0 to 1 we follow a path in composition space that starts at a ternary DPPC/POPC/cholesterol membrane, and that ends at a ternary DPPC/DUPC/cholesterol membrane. The former is homogeneous while the latter is known to separate into coexisting ordered and disordered phases. At intermediate values of $\chi$ the membrane is a quaternary mixture of DPPC, POPC, DUPC, and cholesterol.

\subsubsection{Simulation Parameters and Setup}

Simulations were performed using the GROMACS software package (version 4.6.5)~\cite{Berendsen1995,Pronk2013}. The time step of the simulation was set to be 20fs, which is typical for MARTINI simulations \cite{Marrink2007}. Van der Waals and electrostatic interactions were truncated at 1.2nm, with a smooth decay of the former starting at 0.9nm. Temperature was controlled by a stochastic velocity rescaling thermostat~\cite{gromacs456, Bussi2007}. Pressure was controlled by a semi-isotropic Berendsen barostat~\cite{Berendsen1995} with a reference pressure of 1 bar, which effectively maintains zero surface tension for the membrane.

The construction of the simulated systems proceeded in multiple steps. First, we used the INSANE program~\cite{Wassenaar2015} to build small patches of lipid bilayers consisting of 24 DPPC, 24 POPC, and 24 cholesterol molecules, together with 618 MARTINI solvent particles. This system was equilibrated at 350 K, five-fold replicated in each membrane direction, and then again equilibrated at 350 K. We then replaced a  number of randomly selected POPC molecule with DUPC to obtain starting structures for our systems at different $\chi$ values. The resulting structures contained 1800 lipids and 14540 solvent particles, and had a side length of approximately 40 nm. They served as starting points for production runs at 298 K.

\subsection{\label{subsec:2}Radial Distribution Function}
We define the partial molecular density field of lipid species $\alpha$ for each leaflet as:
\begin{equation}\label{eqn:density}
	\rho_{\alpha}\left(\vek{r}\right) = \sum_{i=1}^{N_{\alpha}}\delta\left(\vek{r}-\vek{r}_i\right)
\end{equation}
where $\vek{r}_i$ is the $(x,y)$ projection of the center of mass position between the two glycerol-ester beads of the $i$-th molecule of lipid species $\alpha$, and $N_\alpha$ is the total number of lipid $\alpha$ in the leaflet. We use the MDAnalysis software package to identify the two leaflets of the bilayer~\cite{Gowers2016}. For cholesterol molecules, due to their tendency to reside in, and sometimes flip-flop between, the two leaflets, the assignment is based on the molecular orientation defined by the vector pointing from the center of geometry of the molecule to the hydrophilic hydroxyl group.

We calculate the partial radial distribution function (RDF), also known as the partial pair correlation function, between lipid species $\alpha$ and $\beta$ 
according to \cite{HansenMcDonald2013,Das2003}
\begin{equation}
	g_{\alpha\beta}\left(\vek{r}\right) = \ffrac{N}{\rho N_{\alpha}N_{\beta}}\left\langle\sum_{i=1}^{N_{\alpha}} \sum_{j=1}^{N_{\beta}}\ '\delta\left(\vek{r} - \vek{r}_{ij}\right)\right\rangle \label{eq:partialgofr}
\end{equation}
where $\vek{r}_{ij}$ is the $(x,y)$ projection of the inter-molecular distance between the center of mass positions of the $i$th and the $j$th lipid molecules, and the prime symbol indicates that the $i=j$ term is excluded if $\alpha = \beta$. $N_\alpha$, $N_\beta$ are the total number of molecules for lipid type $\alpha$ and $\beta$ respectively, $N$ is the total number of molecules in the leaflet, and $\rho$ represents the average lipid density over the entire leaflet. Angular brackets denote the equilibrium average. The function \eqref{eq:partialgofr} was then radially averaged to obtain $g_{\alpha\beta}(r)$.

In our density function definition we follow the convention that the lipid position $\vek{r}_i$ is the center of mass between its two glycerol-ester beads~\cite{Rosetti2012}, projected onto the $(x,y)$ plane. Other definitions of molecular position will yield slightly different pair correlation functions; however these small variations are insignificant for the purpose of this study.

\subsection{\label{subsec:3}Structure Factor}
The structure factor provides information about the spatial organization of the membrane, and can be obtained from neutron or X-ray scattering experiments. Here we assume that each molecule can be considered as a point scatterer at position $\vek{r}_i$. While this is a gross approximation it is justified because we are only interested in large-scale structure where detailed information from single molecules becomes negligible. We define the partial structure factor as~\cite{HansenMcDonald2013}
\begin{equation}\label{eqn:sk_def01}
	S_{\alpha\beta}(\vek{k}) = \left\langle\ffrac{1}{N}\tilde{\rho}_{\alpha}(\vek{k}) \tilde{\rho}_{\beta}(-\vek{k})\right\rangle 
\end{equation}
where 
\begin{equation}\label{eqn:rhok_def01}
	\tilde{\rho}_{\alpha} (\vek{k}) = \int \dif \vek{r} \, \eexp{i \vek{k} \vek{r}} \rho_{\alpha}(\vek{r}) 
\end{equation}
is the Fourier transform of the partial lipid density function from equation \eqref{eqn:density}. The structure factor and radial distribution function are related by ~\cite{HansenMcDonald2013}
\begin{equation}\label{eqn:sk_def02}
	S_{\alpha\beta}(\vek{k}) = x_{\alpha}(\delta_{\alpha\beta}+ x_{\beta})\rho\int g_{\alpha\beta}(r)\eexp{-i\vek{k}\cdot\vek{r}}\mathrm{d}\vek{r} 
\end{equation}
where $x_\alpha$ and $x_\beta$ are the mole fractions of lipid species $\alpha$ and $\beta$ respectively, and $\delta_{\alpha\beta}$ is the Kronecker delta.

\subsection{\label{subsec:4}Local Density Distribution (LDD)}
We define the local density distribution (LDD) as the probability distribution function of the instantaneous density of lipid species $\alpha$ in a region linear dimension $w$:
\begin{equation}
	P(\rho_{\alpha}) = \frac{1}{A} \int \dif \vek{r} \left\langle \delta \left( \hat{\rho}_{\alpha} (\vek{r} ) - \rho_{\alpha} \right) \right\rangle \label{eq:LDDdefinition}
\end{equation}
where 
\begin{equation}
	\hat{\rho}_{\alpha} (\vek{r} )  = \frac{1}{w^2}\int _ {(w(\vek{r}),w(\vek{r}))} \dif \vek{r}'\rho_{\alpha}\left(\vek{r}'\right) 
\end{equation}
is the local partial lipid density averaged over a square observation window with side length $w$ positioned at $\vek{r}$ on the membrane. $A$ is the total area of the membrane. In practice \eqref{eq:LDDdefinition} is calculated by sweeping the observation window laterally over each leaflet in every simulation frame.

To help us understand the LDD profile we develop a simple continuum model as described in the Supporting Information (SI). We show that the characteristics of the LDD profile depends both on the degree of lipid segregation and on the observation window size, $w$. We provide detailed analysis of this dependence in the Results section.

\subsection{\label{subsec:5}Local Composition Distribution (LCD)}
Similar to the LDD, we also calculate the local composition distribution (LCD) function with the same observation window algorithm to analyze lipid composition on the bilayer system. Just like in the calculation of LDD, we employ observation windows at various sizes to sample the local composition of each individual lipid species within the observation window, and the window then sweeps both leaflets separately for the entire bilayer to acquire statistical information on the local composition throughout the membrane. We compute the multi-dimensional probability distribution:
\begin{equation}
	 P(\vek{\Gamma}) =\ffrac{1}{A}  \int \dif \vek{r}\left\langle \prod_{\alpha} \,  \delta \left( \hat{\Gamma}_{\alpha} (\vek{r} ) - \Gamma_{\alpha} \right) \right\rangle
\label{eq:LCD}
\end{equation}
where 
\begin{equation}
	\hat{\Gamma}_{\alpha} (\vek{r} )  = \ffrac{N_{\alpha, w}(\vek{r}) }{N_{tot, w}(\vek{r} ) } 
\label{eq:LCDsample}
\end{equation}
is the local partial lipid composition of lipid species $\alpha$ within an observation window of size $w$ centered at position $\vek{r}$, and $N_{\alpha, w}(\vek{r}) $ and $N_{tot, w}(\vek{r} )$ are the number of lipid $\alpha$ molecules and total number of all lipid molecules within the observation window centered at $\vek{r}$, respectively. $\vek{\Gamma}$ is the three- or four-dimensional vector of local compositions for each lipid species, and $A$ is the total area of the membrane.

\subsection{\label{subsec:6}Gaussian Mixture Model}
To analyze the multi-dimensional LCD profile for our four-component lipid mixture and quantify the onset of phase separation, we use a Gaussian Mixture Model as implemented in the Python Scikit-learn machine learning package developed by Pedregosa \textit{et.} \textit{al.} \cite{pedregosa2011} to identify the nature of the distribution of lipid composition. Inspired by the continuum bilayer model (described in the SI), we perform three population Gaussian cluster analysis to the multi-dimensional local lipid composition distribution in order to identify the lipid-rich and lipid-poor phase and their corresponding compositions. The Bayesian Information Criterion (BIC) for Gaussian mixture models was used to ascertain whether a three-population fit or a one-population fit is more appropriate~\cite{Kass1995}, corresponding to phase-separated and homogeneous membranes. The BIC is defined as
\begin{equation}
	\mathrm{BIC} = n \cdot \ln\left(\hat{\sigma_e^2} \right) + k \cdot \ln \left( n \right)
\end{equation}
where $n$ is the the number of data points in the random variable, $k$ is the number of free parameters used in the fit, and $\hat{\sigma_e^2}$ is the error variance, defined as:
\begin{equation}
	\hat{\sigma_e^2} = \ffrac{1}{n} \sum_{i=1}^{n} \left( x_i - \hat{x_i} \right)^2
\end{equation} 
where $x$ is the random variable, or in this case, the local lipid compositions. Details of the Gaussian mixture and BIC analysis are shown in the SI.

\section{\label{sec:results}Results}

\subsection{Domain Formation and Local Density Distribution Depend on Composition}

\begin{figure}[t]
\begin{center}
\includegraphics[width=\columnwidth]{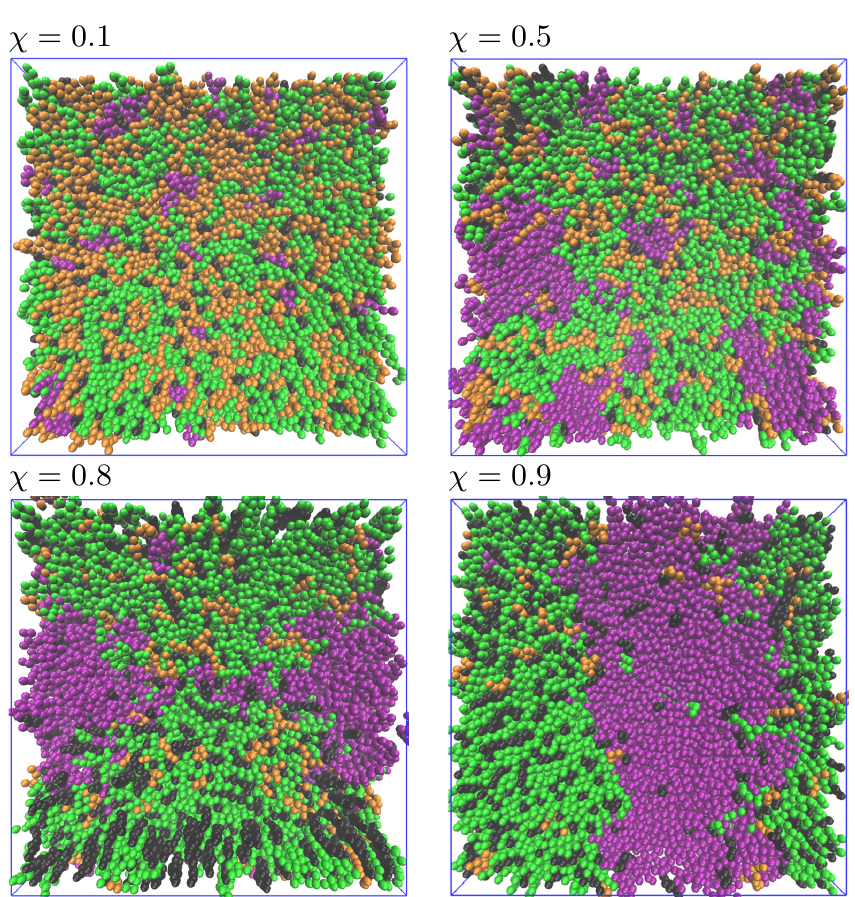}
\end{center}
\caption{\label{fig:bilayers} Simulation snapshots of quaternary lipid bilayers taken 20 $\mu$s after the initial temperature quench with increasing $\chi$ values of $0.1$ (top left), $0.5$ (bottom left), $0.8$ (top right), and $0.9$ (bottom right), with DPPC (green), DUPC (purple), POPC (orange), and cholesterol (black). At low $\chi$ the system exhibits no apparent long range order. As $\chi$ increases we start to observe small clustering of domains enriched in DUPC. Further increasing the DUPC content induces phase separation into DUPC-rich and DUPC-poor regions.}
\end{figure}

Final configurations of four (out of eleven) MD trajectories at a simulation time of $20 \mu$s are shown in Figure~\ref{fig:bilayers}. At low $\chi$ value the system exhibits no apparent long range order. As $\chi$ increases we start to observe domains enriched in DUPC. Further increasing the DUPC composition causes formation of long-lasting, large scale DUPC-rich and DUPC-poor regions. Due to the geometry of the rectangular simulation box together with the use of periodic boundary conditions the shape of the coexisting domains can be either circular or stripe, depending on the area fraction of the domains. 

To quantify the bilayer heterogeneity we calculate the local lipid density distribution (LDD) function of the four lipid species. We calculate the number density of each lipid type within an observation window that is smaller than the simulation box size. Statistics on the local lipid density were collected at different locations of the observation window throughout the entire membrane and over the last $5\mu $s of the simulation trajectories. Intuitively one would expect the distribution to be unimodal if the membrane is homogeneous, i.e., the bilayer system has the same composition everywhere. On the other hand, if the bilayer consists of distinct regions where a lipid species is either enriched or depleted, we expect a bimodal distribution of partial lipid densities indicating two distinct populations in the lipid density distribution. 
To make this intuitive picture quantitative we devised a simple continuum model (shown in the {SI}) to capture the spatially heterogeneous nature of the membrane and to analyze it by deriving an analytical form of the LDD for such heterogeneous systems. The results are then used to analyze the lipid density distribution from bilayer MD simulations.

\begin{figure} [t]
\includegraphics[width=0.45\textwidth]{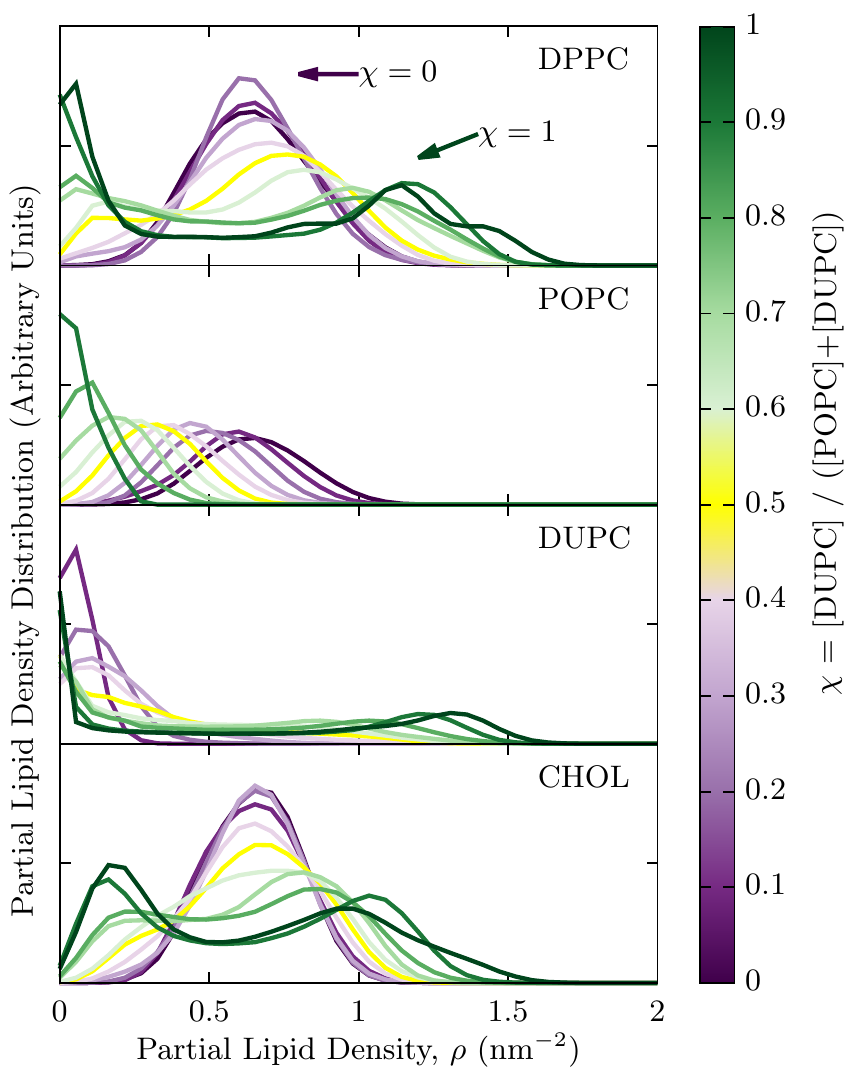}
\caption{\label{fig:histgrid_wsize20} Local density population distributions calculated using the sampling window algorithm (with window size $4 \mathrm{nm}\times 4 \mathrm{nm}$) of all simulated bilayer systems for all lipid types. The colored spectrum and the arrows indicate the value of $\chi$ of the system, from high (green) to low (purple), with the half-way point singled out (yellow). For compositions with low content of DUPC ($\chi < 0.5$), unimodal distributions of partial lipid density are observed for all lipid types, while the peak for DUPC is less Gaussian-like due to insufficient sampling of DUPC molecules at very low DUPC composition. For compositions with high $\chi$ values bimodal distributions of lipid density are found for DPPC, DUPC, and cholesterol. For POPC the density distribution remains unimodal while being shifted to the low density end due to decreases global POPC density, and the peak also gets distorted away from its Gaussian shape due to decreased sample size.
}
\end{figure}

\begin{figure} [t]
\includegraphics[width=\columnwidth]{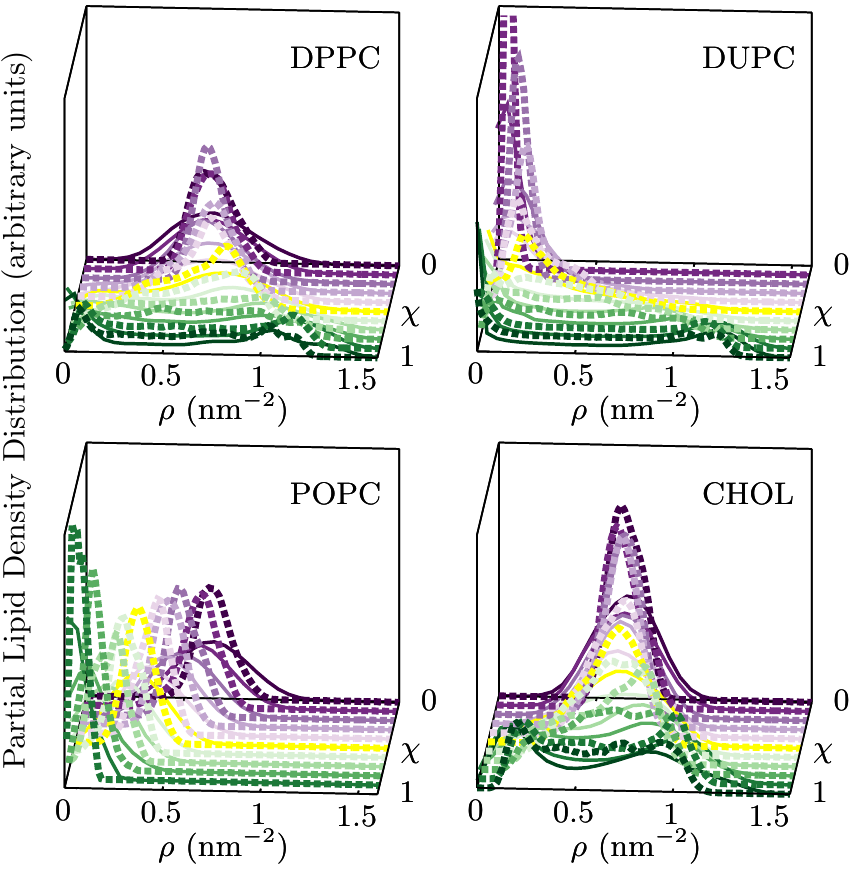}
\caption{\label{fig:histgrid_wsize2040} LDD calculations for all lipid species at various compositions with different window sizes, one of which was shown in {Figure \ref{fig:histgrid_wsize20}} with  $4 \mathrm{nm} \times 4 \mathrm{nm}$ observation window (solid lines), and the other one with bigger observation window of $8 \mathrm{nm} \times 8 \mathrm{nm}$ (dashed lines). The color scheme is based on composition variable $\chi$ and follows the same spectrum in {Figure \ref{fig:histgrid_wsize20}}. At large $\chi$ values both LDD profiles show bimodal distributions of lipid density for DPPC, DUPC, and cholesterol. It is noted that, for a bigger observation window size the bimodal-to-unimodel transition in the density distribution happens at larger $\chi$ values compared to LDD with smaller observation window. For DPPC, DUPC, and cholesterol LDDs at small $\chi$ values, as well as for POPC LDDs at all compositions, we observe unimodal density distribution, indicating homogenous distribution at the observed length scales. The calculations with bigger observation window show a narrower distribution in local density than those with smaller window at all compositions. }
\end{figure}

Results of the LDD calculation for all four molecular species and 11 studied membrane compositions are shown in Figure~\ref{fig:histgrid_wsize20}. We find unimodal local density distributions of DPPC molecules (top panel) in quaternary systems at low $\chi$ values ($0-0.4$). This shows that no large-scale composition heterogeneity exists at and beyond the length scale of the $4 \mathrm{nm} \times 4 \mathrm{nm}$ window we used. At higher $\chi$ ($> 0.6$), a bimodal pattern starts to emerge, indicating two distinct populations of local lipid densities, indicating the coexistence of DPPC-rich and DPPC-poor regions. We also observe an increase in the difference between the two peaks of the bimodal DPPC local density distribution for higher $\chi$ values, which suggests that the compositions in the two regions become increasingly different as more POPC is replaced by DUPC. It is noticeable that, as the distribution shifts from the unimodal regime to the bimodal regime, the emerging low density peak does not retain the Gaussian-type distribution of the unimodal peak. This is caused by the small number of DPPC molecule in the $L_d$ domain and the small observation window size. In this case the density distribution can no longer be properly described by a Gaussian distribution, but should rather be modeled by a Poisson distribution, both of which are illustrated in a discrete membrane model in the {SI}. No significant difference is observed between the two leaflets and the reported data were averaged across both leaflets. The local density distributions of DUPC molecules show a similar progression from unimodal to bimodal behavior as the composition variable $\chi$ increases, which strongly suggests a similar heterogeneous partitioning of DUPC molecules as in the case of DPPC. In contrast to DPPC, the global density of DUPC changes with $\chi$, therefore the center of the peak shifts as $\chi$ changes, and the histograms look qualitatively different between DPPC and DUPC as a result.

For POPC molecules no bimodal distribution is observed over the entire $\chi$ range. Due to the change in global POPC composition as we traverse the $\chi$ axis, the POPC composition peak shifts, similar to the case of DUPC. However the local density distribution of POPC remains unimodal, which suggests that the POPC lipid does not preferentially partition into any of the coexisting phases in a significant way. When the global POPC density is low at small $\chi$ values we again see deviations form Gaussian behavior as discussed previously.
 
The LDD profiles of cholesterol are very similar to those of DPPC. At small $\chi$ values ($\chi < 0.5$) the distributions show unimodal, Gaussian-like features. As $\chi$ increases the distribution becomes bimodal. The average cholesterol density, or the mean of the distribution, is independent of $\chi$ because the overall cholesterol  composition is kept constant. The separation between the cholesterol-rich domain density and the cholesterol-depleted domain density is not as large as that of the two phospholipids, indicating that the preferential partitioning of cholesterol between the two regions is not as strong as that of DPPC and DUPC. Another feature of the cholesterol LDD that is consistent with this observation is that the low density peak cholesterol is much better resolved and retains a Gaussian-type line shape around its maximum. Both observations indicate that the cholesterol content in the two coexisting regions is more similar than the content of the two phospholipids. 

A closer investigation of the LDD profile for both the bilayer simulations and the simple continuum model (shown in the SI) reveals that lipid density distribution profiles depend strongly on the scale of potential heterogeneities and the observation window size by which this density profile is obtained. The same LDD calculation was performed with a different observation window size of $8 \mathrm{nm} \times 8 \mathrm{nm}$. Results are shown alongside those obtained with a smaller window in Figure~\ref{fig:histgrid_wsize2040}. By using a larger sampling window size we observe that if a lipid composition show unimodal density distributions with a smaller observation window, the larger window still indicates spatial homogeneity. It is noticeable that the distribution gets narrower as the window gets larger, which is a consequence of the central limit theorem. More interestingly, for some systems where the LDD profile for the smaller window indicates two coexisting phases, that of the larger window shows no strong evidence of a bimodal distribution. Further analysis suggests that the two peaks in a bimodal density distribution start to merge when the observation window size exceeds a $\chi$-dependent threshold value. Comparison with the simple continuum model described in the SI shows that this threshold size is an indicator for the length scale of the heterogeneity.

In the following two sections we will first focus on quantifying the difference between coexisting phases and domains in order to understand the nature of the relationship between membrane composition and membrane morphology. Then we will elucidate the dependence of LDD on observation window size, and its implication on identifying the nature, as well as quantifying the degree, of lipid spatial heterogeneity in bilayers systems.

\subsection{Using Density Correlation Functions to Identify the Onset of Phase Separation }

\begin{figure} [t]
\includegraphics[width=\columnwidth]{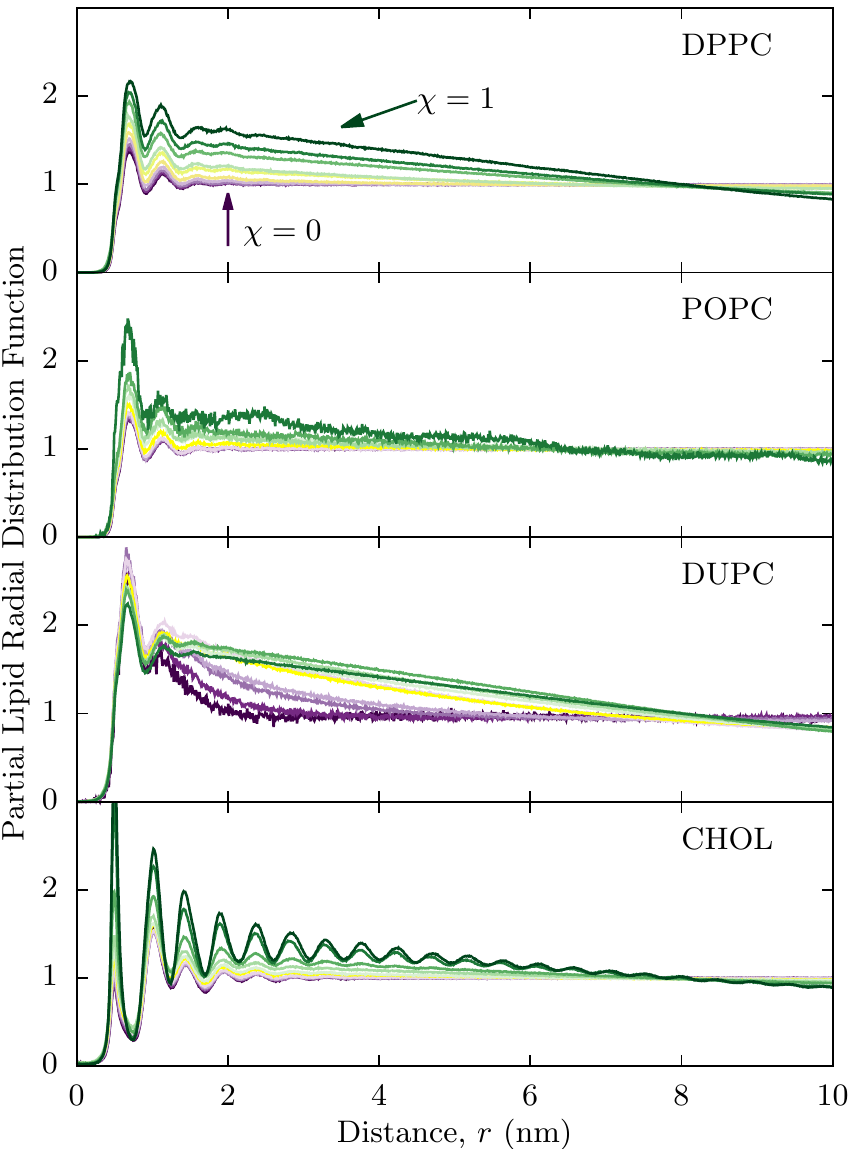}
\caption{\label{fig:rdfs_avg} Partial radial distribution functions of lipids in bilayer systems for different molecule types at various compositions. The color scheme is the same as in Fig.~\ref{fig:histgrid_wsize20}, ranging from low $\chi$ (purple) to high $\chi$ (green) value, with $\chi = 0.5$ (yellow) being the halfway point. At low $\chi$ values the RDF approaches unity, indicating composition homogeneity over large distances. At high $\chi$ values the RDFs for DPPC, DUPC, and cholesterol show a characteristic long-range linear decay, a hallmark of phase separation, while the RDF of POPC converges to unity at long range for all $\chi$ values. 
}
\end{figure}

Pair correlation or radial distribution functions (RDFs) provide a different route to studying membrane heterogeneity. Figure~\ref{fig:rdfs_avg} shows results obtained from our coarse-grained simulations. 

At short distances the RDF provides information on molecular packing. It is noticeable that in this regime the RDF varies systematically with membrane composition. For example, in the case of DPPC the height of the nearest-neighbor peak increases with $\chi$.
This behavior is expected for a system that undergoes a transition from a homogeneous to a phase-separated regime, where in the latter the DPPC molecules are concentrated in one domain and depleted in the other. 
The RDF of cholesterol shows long-range order at high $\chi$ values, indicating nearly crystalline order. This is a known artifact of the MARTINI coarse-grained model that has been rectified in recent versions of this force field~\cite{Ingolfsson2014}.

At large distances the RDFs of DPPC, DUPC and cholesterol exhibit a long-ranged, nearly linear decay to values less than unity at compositions rich in DUPC. This indicates that the system cannot be homogeneous at those large $\chi$ values, because homogeneity invariably translates into flat pair correlation functions at large distances, indicating the finite range of density correlations in bulk fluids. The fact that such uniformity is not observed even at length scales comparable to the size of the system is a strong indicator for thermodynamic phase separation. 

To obtain further insight from these RDFs we analyze in the Supplemental Information a simple continuum model of two coexisting phases, both with spherical and stripe domains. We derive analytical expressions for the pair correlation function, which exhibits a nearly linear, slowly decaying behavior at large length scales, similar to those shown in Figure~\ref{fig:rdfs_avg}. The analysis of the continuum model shows that the slope of the linear part of the RDF is related to the difference $\Delta \rho$ in density of a lipid species between the two coexisting regions. With this analysis at hand we can extract this important quantity for the quaternary membrane studied in our simulations.

\begin{figure}[tb]
\includegraphics[width=\columnwidth]{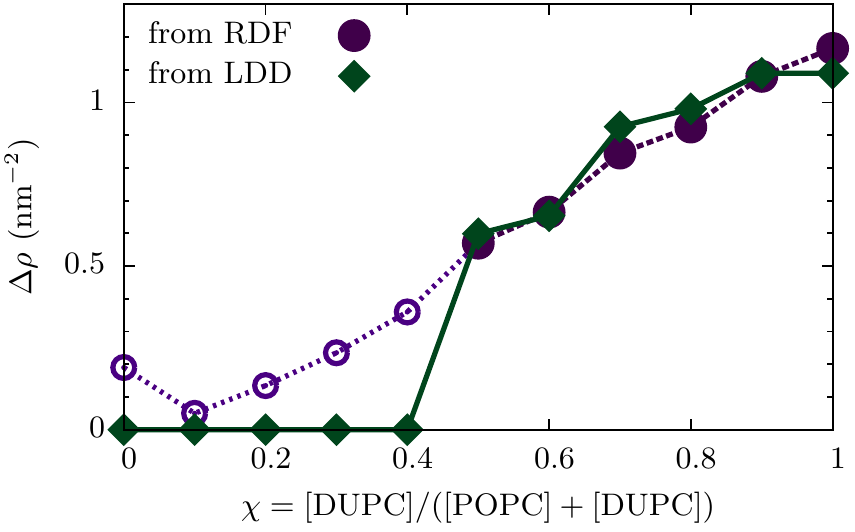}
\caption{\label{fig:delrho_dppc02} Density difference of DPPC between coexisting phases, calculated using the slope of the radial distribution function (Fig.~\ref{fig:rdfs_avg}) and from the local density distribution function (Fig.~\ref{fig:histgrid_wsize20}).
At large values of $\chi$ both the RDF and the LDD results show significant density contrast between the DPPC-enriched and the DPPC-depleted region. This magnitude of this contrast depends on composition. At low values of $\chi$ the RDF approach becomes less reliable (open symbols) due to difficulties in identifying and fitting the linear region of the density correlation function.}
\end{figure}

At each simulated composition, the partial RDFs were calculated, and the long-ranged linear portions were identified and fit to straight lines. Using the relationship obtained for the continuum model (equation (S11)) we calculated the density contrast. Figure~\ref{fig:delrho_dppc02} shows the results of this approach for DPPC, and compares them to the difference in peak positions of the LDD profiles discussed in the previous section. Both approaches find significant density contrast between the two phases at large values of $\chi$, and their estimates of $\Delta \rho$ agree quantitatively. For $\chi \leq 0.4$, however, the two approaches give conflicting results: while the LDD for these compositions is unimodal, which points toward a single homogeneous phase, the linear fit to the pair correlation function yields a negative slope, which gives rise to a small but non-zero estimates of the density contrast. In this regime the RDF approach becomes unreliable, because the slope of the linear fit, and therefore the estimate for $\Delta \rho$, become sensitive to the range over which the RDF is fitted. Furthermore, this method assumes that phase-separated domains, if they exist, form two rectangular stripes in the system. While this is the case for separated systems with high interface tension, this assumption breaks down if the interface tension becomes small and comparable to thermal fluctuations. The LDD method does not require such assumptions, we believe it is therefore the better approach to establish the presence of coexisting domains and the density contrast between them.

\begin{figure} [h]
\includegraphics[width=\columnwidth]{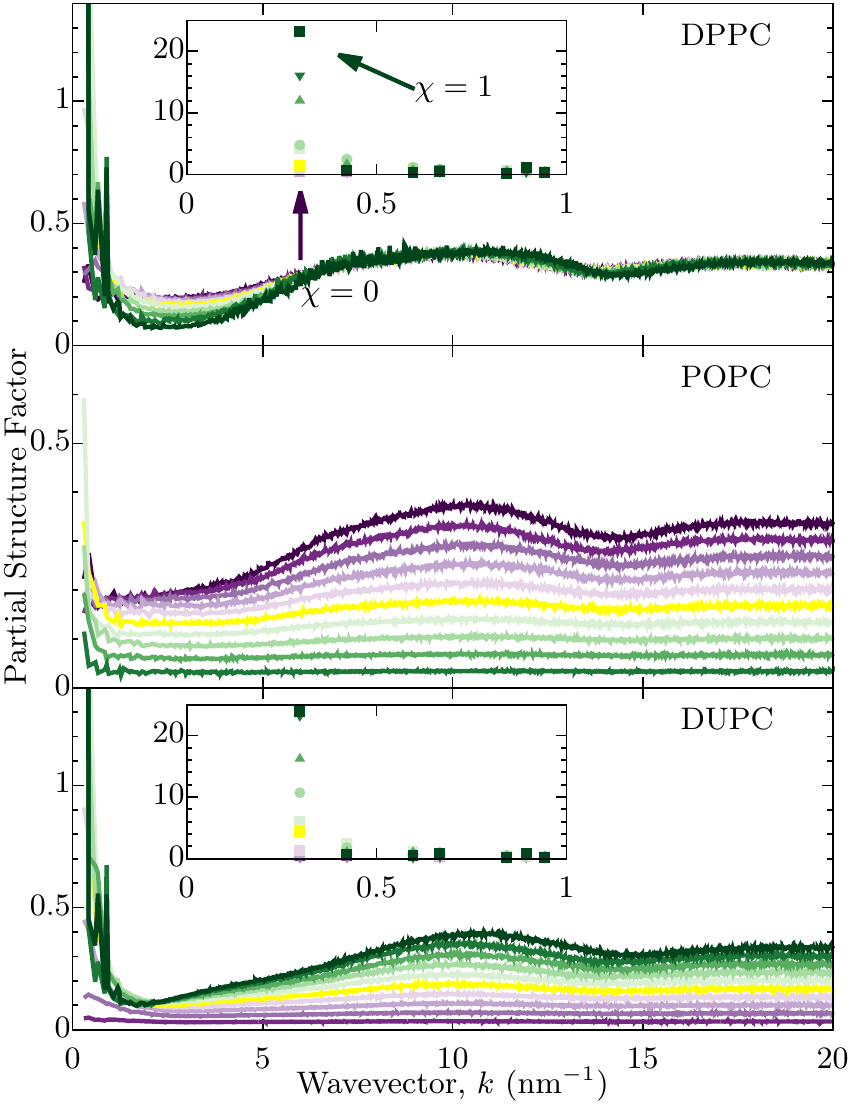}
\caption{\label{fig:sfacs_comb_all02} Radially averaged 2-d partial structure factors for all studied compositions and lipid species. The color scheme is the same as in Fig.~\ref{fig:histgrid_wsize20}. For lipid species with high concentrations (DPPC, DUPC at high $\chi$, and POPC at low $\chi$) we observe a broad peak at $\approx 10\ \mathrm{nm}^{-1}$, which originates from nearest neighbor packing. At very large $k$ values the structure factor converges to the global mole fraction of the given lipid, without accounting for cholesterol molecules. Large values of the structure factor at low $k$ indicate significant density fluctuations on the scale of the system. Insets show that the magnitude of such density fluctuations is significantly larger for DPPC and DUPC than for POPC at high $\chi$, while they are comparable at low values of $\chi$.
}
\end{figure}

\subsection{Analysis of Structure Factors}

A complementary way to detect and characterize lipid heterogeneity is the analysis of the partial structure factors $S_{\alpha\beta}(\vek{k})$, defined in equation \eqref{eqn:sk_def01}. While containing the same information as the pair correlation function, the structure factor is of interest for multiple reasons. First, it is directly related to data obtained in neutron or x-ray scattering experiments~\cite{Heberle2013, Armstrong2013}. Second, the structure factor can be used to distinguish between unstructured fluids, structured fluids, and modulated phases. Recent work has suggested that seemingly homogenous bilayers might in fact be microemulsions that are ordered on length scales not discernible in optical experiments~\cite{Schick2012, Shlomovitz2013, Shlomovitz2014}. The hallmark feature of this microemulsion phase is a peak in the structure factor at non-zero wavevector on length scales larger than those related to molecular packing.

Figure~\ref{fig:sfacs_comb_all02} shows partial structure factors for the three phospholipids in all studied systems. They show significant changes in bilayer structure as the composition is varied. Their normalization is such that they converge at large wavevectors to a lipid species' mole fraction in the membrane. As the parameter $\chi$ changes from 0 to 1 this mole fraction changes for POPC and DUPC but not for DPPC, which is why the partial structure factors of the latter overlap while those of the first two lipids are separated by a constant offset.

We are primarily interested in the low wavevector (large length scale) regime. Both the DPPC and the DUPC partial structure factors exhibit 5-10 fold higher low-wavevector intensities at large vs. low $\chi$ values, indicating long range order of density fluctuations for phase-separated systems and large scale segregation of DPPC and DUPC lipid molecules. The POPC structure factor shows no significant variation in the low wavevector regime with increasing $\chi$. This suggests, as one would expect from the LDD and RDF results, that POPC does not participate strongly in the segregation of lipids, while DPPC and DUPC show strong partitioning into heterogeneous domains.

The nature of this apparent heterogeneity cannot be identified from the analysis of partial structure factors alone. A sharp increase of $S(k)$ as $k \rightarrow 0$ is consistent with two coexisting thermodynamic phases, separated over the length scale of the entire system. Alternatively, a peak in $S(k)$ at small but non-zero wavevector is consistent with a single, structured phase such as a microemulsion~\cite{Schick2012,Shlomovitz2013,Shlomovitz2014}. Our simulations cannot distinguish between these two scenarios, which would require additional data at lower wavevectors to test whether the observed behavior of $S(k)$ is part of a monotonic increase at low wavevectors or the high-$k$ flank of a peak in the structure factor. This data can only be provided by simulations spanning much larger system sizes, which are currently unfeasible. 

Combining results from LDD, RDF, and structure factor calculations, along with a model-based analysis of lipid density distributions, we have established that both the length scale and the lipid composition of heterogeneous lipid domains depend on the membrane composition. Although the phase behavior at the near-ternary end points of $\chi$ has been well studied by experiments and simulations (including this work), the nature of this heterogeneity remains unclear. As indicated by the high intensity at low wavevectors in the partial structure factors, the size of heterogeneous domains are apparently often bound by the simulation box size, which does not allow us to distinguish between multiple segregated phases and a single structured fluid. While performing simulations on significantly larger systems is out of reach, we can obtain related information by varying the size of the observation window inherent to the LDD analysis method while keeping the system size fixed. 

\subsection{\label{subsec:LDDobservationsizedependence}Dependence of Local Density Distributions on Observation Window Size}

\begin{figure}[t]
\includegraphics[width=\columnwidth]{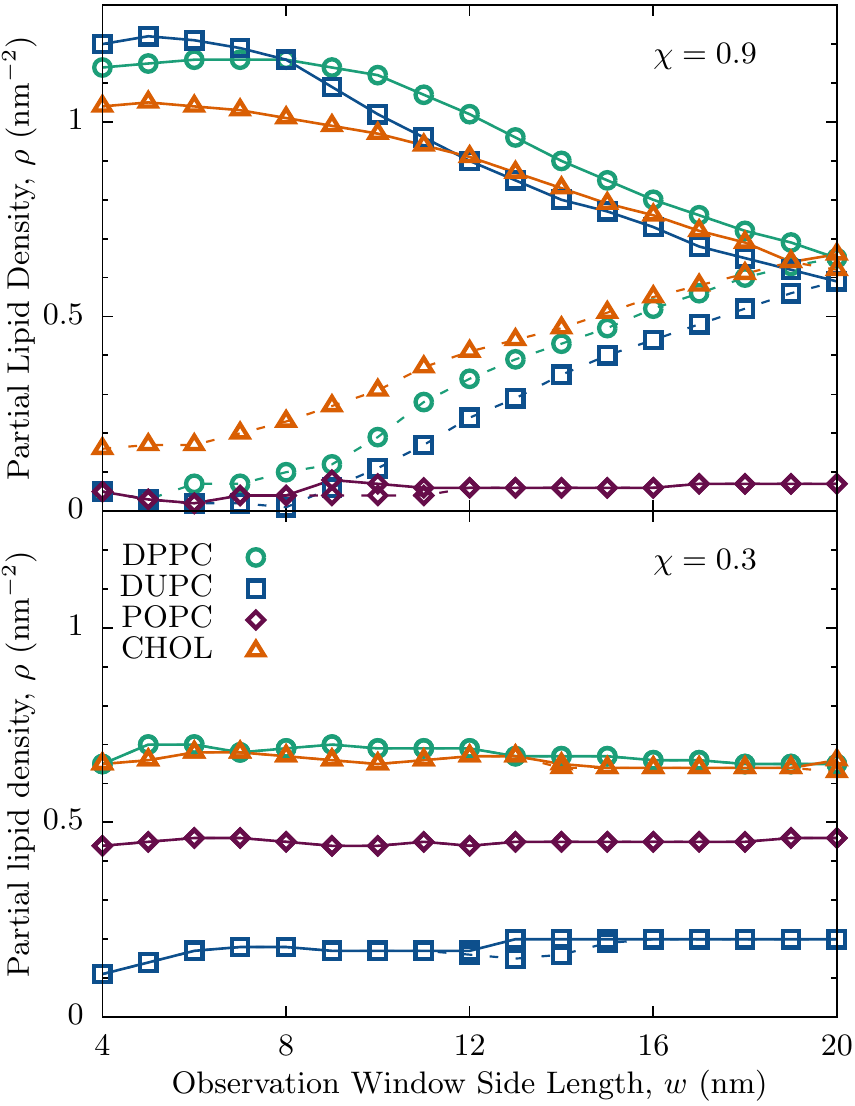}
\caption{\label{fig:histpeak_bycompochi09} 
Peak positions in the local density distribution (LDD) function for various lipid species as a function of observation window size $w$. In a segregated system ($\chi = 0.9$, top panel) the LDDs for DPPC, DUPC and cholesterol have two local maxima corresponding to the densities in distinct enriched or depleted domains, whereas the LDD of POPC only has a single peak independent of $w$. In a homogenous system ($\chi = 0.3$, bottom panel) the LDDs of all species have only a single peak independent of observation window size.
}
\end{figure}

As previously discussed and illustrated in Figure~\ref{fig:histgrid_wsize2040}, the LDD depends on the size $w$ of the observation window chosen for the analysis. This dependence can be used to gain further insight into the nature of lipid heterogeneity in mixed bilayer systems.
For example, Figure~\ref{fig:histpeak_bycompochi09} shows the position of local maxima in the LDDs of the four lipid types as a function of $w$ for two different membrane compositions.

For bilayers rich in DUPC ($\chi = 0.9$, top panel) we find that when the observation window is sufficiently small ($w < 8$ nm) two distinct populations of local lipid densities are observed for DPPC, DUPC and cholesterol, indicating a separation of the bilayer into regions enriched and depleted of those species. POPC, on the other hand, does not participate in this separation. As predicted by the analytical form of the LDD in a simple continuum model (shown in the SI), when the observation window size goes beyond the length scale of the heterogeneity the spacing between the two peaks decreases, as the calculation now fails to sample the domain bulk lipid density due to the large observation window size. For the strongly phase-segregated system the rich-phase density only becomes identical to the poor-phase density when the observation window is as big as the entire system. This suggests that the length scale of lipid density heterogeneity scales with the simulation box size, which is indicative of thermodynamic phase separation.

For bilayers with low DUPC content ($\chi = 0.3$, bottom panel) we find that the LDDs of each lipid species exhibits only a single maximum whose position is nearly independent of the observation window size $w$. These results suggest that the bilayer is laterally homogeneous over the entire range of length scales that we considered. It is noticeable that when the observation window gets very small (close to 4 nm) some lipids, especially DUPC, show a slightly lower-than-average local density. This is most likely due to the statistics of low numbers in the limit of low overall DUPC density and small observations windows.  We attribute the brief appearance of a DUPC-poor peak around $w = 14$ nm to the uncertainty inherent in identifying local maxima in the noisy LDD that contains statistical uncertainty.

\begin{figure}[tb]
\includegraphics[width=\columnwidth]{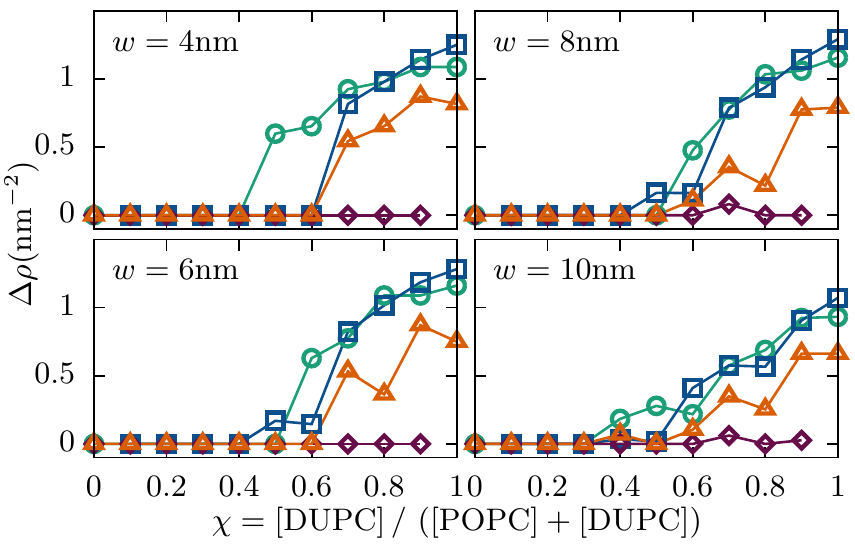}
\caption{\label{fig:histpeak_delrho} The density contrast between enriched and depleted local lipid densities as a function of observation window size. Symbols and colors are as in Fig.~\ref{fig:histpeak_bycompochi09}. }
\end{figure}

To further study the relationship between the degree of lipid heterogeneity and observation window size we compute the lipid density contrast $\Delta \rho$ between the enriched and the depleted region for each lipid type as a function of global composition $\chi$. These calculations are similar to the LDD calculation shown in Figure~\ref{fig:delrho_dppc02} for DPPC, but are now done for all lipid types at various compositions and observation window sizes. Results are shown in Figure~\ref{fig:histpeak_delrho}. First, we see that for reasonably small window sizes ($w < 11$ nm, about half of the simulation box size) the system shows no sign of heterogeneity for compositions with small $\chi$, as the lipid density contrast between the two regimes remains zero. Second, for large $\chi$ the system shows strong segregation as shown by large differences in local lipid densities. These values correspond to the end points of the tie-line projected onto the axis of each lipid species. It is noticeable that for smaller observation window sizes ($w$ = 4, 6 and 8 nm) the density contrast for the same lipid type converges to the same value as $\chi$ approaches 1 consistently across different window sizes. This indicates that these observation windows are small enough to capture the bulk phase densities. For $w$ = 10 nm, however, the maximum contrast at $\chi = 1$ starts to decrease compared to the value from smaller windows, suggesting that the observation window has grown beyond the length scale of heterogeneity of the system.

Our analysis of the local density distribution functions shows that in order to obtain accurate estimates of bulk lipid densities in coexisting domains one has to use an observation window size that is significantly larger than the size of an individual lipid, but that cannot exceed the length scale of the heterogeneity. In case of thermodynamic phase coexistence the latter is determined by the size of the system. Characteristic changes in the LDD as the size of the observation window is varied can be used to demonstrate phase coexistence, and to distinguish it from other mechanisms that may lead to spatial heterogeneity.

\subsection{Obtaining composition phase diagrams using Local Composition Distributions}

\begin{figure} [t]
\includegraphics[width=\columnwidth]{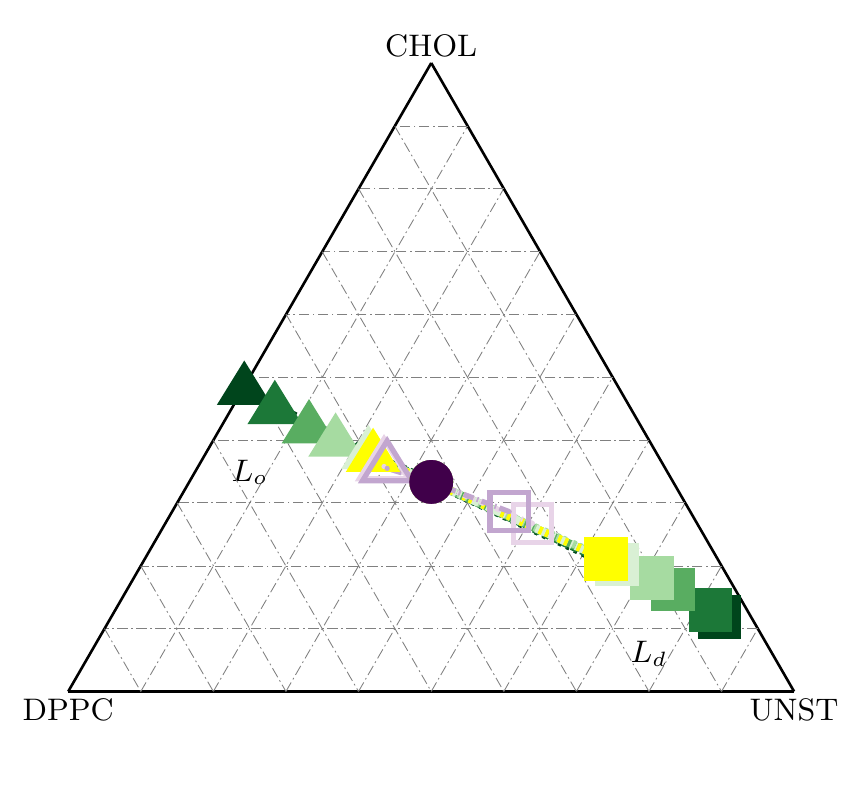}
\caption{\label{fig:phasediag_w50} Multi-component lipid phase diagram of the quaternary lipid mixture as obtained from the Gaussian Mixture Model analysis. UNST stands for the sum of unsaturated POPC and DUPC lipids. Colors are as in Fig.~\ref{fig:histgrid_wsize20}. At large $\chi$ values the Bayesian Information Criterion (BIC) indicates strong evidence for two distinct populations of lipid species (solid symbols), and the system exhibits coexistence of a liquid-ordered ($L_o$) phase rich in DPPC and cholesterol (triangles) and a liquid-disordered ($L_d$) phase rich in unsaturated lipids (squares). At intermediate $\chi$ values the evidence for phase separation is very weak (hollow symbols). At low $\chi$ the LCD shows a unimodal distribution of lipid compositions and the BIC supports the identification of a single, uniform phase (solid circle).
}
\end{figure}

The mixing behavior of multicomponent systems is best illustrated in composition phase diagrams. Compositions (or mole fractions) are generally preferred over molecular densities due to the built-in constraint that the mole fractions of all species must add to unity, thereby reducing the dimensionality from four to three. We therefore adapt our algorithm to calculate the LDD to compute local composition distribution (LCD) functions, defined by \eqref{eq:LCD}. From our simulation data we obtain a large number of samples of local lipid compositions. As described in the Methods section and in the Supplemental Information, these samples are then analyzed by fitting to a Gaussian Mixture Model with either one or three population centers, and the Bayesian Information Criterion (BIC) is used to determine which model best describes the data. This approach allows us to distinguish a homogeneous from a heterogeneous system in a statistically meaningful way.

To simplify the analysis and graphical representation of the mixing behavior we chose to combine the two unsaturated lipids POPC and DUPC into a single component, denoted UNST, thereby further reducing the dimensionality from three to two. This simplification is justified by the LDD calculations shown in Figure~\ref{fig:histpeak_bycompochi09}, because POPC does not show significant separation across all studied observation length scales. Therefore no information is lost by combining POPC with any other lipid species, or by leaving it out of the LCD calculation entirely. We chose to combine POPC with DUPC because the total number of these two lipids is conserved across all compositions considered in this study.

Figure~\ref{fig:phasediag_w50} shows the ternary DPPC:UNST:Cholesterol phase diagram obtained from the LCD analysis. At $\chi \geq 0.5$ the BIC indicates signifiant evidence for a multi-population model of local compositions, which is consistent with a phase-separated system. As illustrated in the SI, the three populations correspond to the compositions of two coexisting bulk phases and an additional broad band that connects them, which stems from samples in which the observation window contains the interface between two domains. One phase is rich in DPPC and cholesterol, while the other is rich in unsaturated lipids. These phases form the end-points of tie-lines that span the coexistence region in the phase diagram. We identify them with the liquid-ordered ($L_o$) and liquid-disordered ($L_d$) phase, respectively.

For $0.3 \leq \chi \leq 0.4$ there is still some evidence for multiple populations in composition space but that evidence is very weak. For $\chi \leq 0.2$ the BIC indicates that a three-population model overfits the data, and that a one-population model more accurately represents the data. To these systems we therefore assign only a single phase, with a composition equal to the global mole fractions of lipids.

\begin{figure} [t]
\includegraphics[width=\columnwidth]{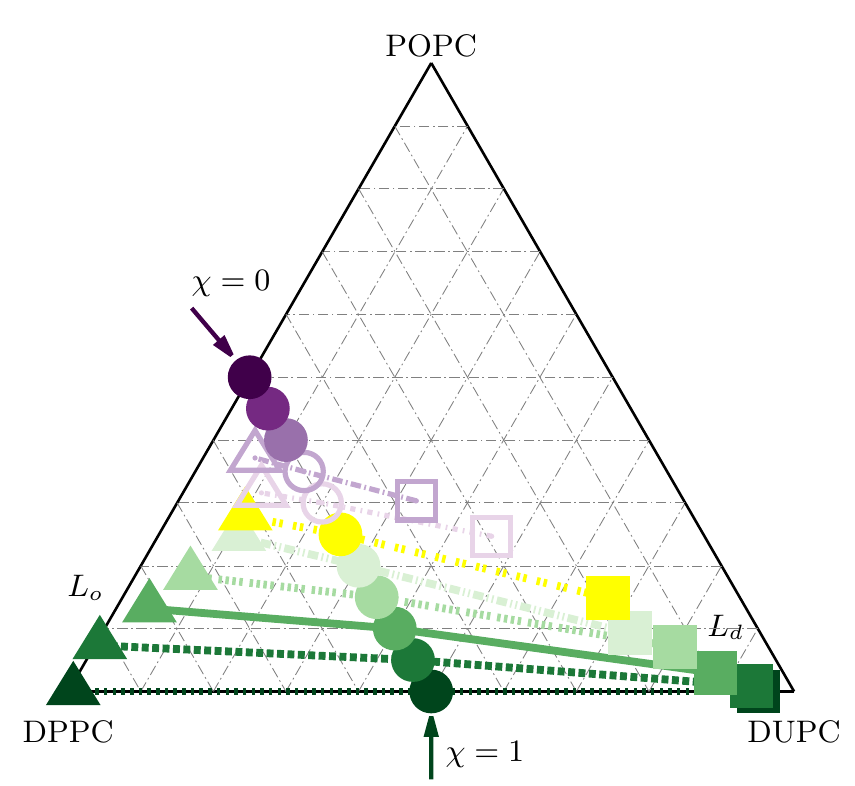}
\caption{\label{fig:phasediag_nochol_nocent_w50} Multi-component phase diagram obtained using a four-dimensional Gaussian Mixture model, and neglecting the cholesterol component for ease of graphical illustration. Colors and symbols are as in Fig.~\ref{fig:phasediag_w50}. In this representation the $\chi=0$ system lies on the DPPC/POPC axis, while $\chi=1$ corresponds to a point on the DPPC/DUPC axis. The latter shows strong separation into a liquid-ordered and a liquid-disordered phase. The contrast between these phases decreases with decreasing $\chi$, and eventually vanishes as the system becomes homogeneous. The tie-lines spanning the coexistence region are slightly titled, indicating a weak preference of POPC for the ordered phase.
}
\end{figure}

To further study the role of POPC we repeat the Gaussian mixture model analysis on the four-component compositions of all lipid species. Figure~\ref{fig:phasediag_nochol_nocent_w50} shows a different rendering of the quaternary phase diagram obtained by projection onto the DPPC-POPC-DUPC plane, thereby ignoring the contribution of cholesterol. As shown previously, cholesterol  preferentially partitions with DPPC into the $L_o$ phase, and by leaving cholesterol out of the analysis we therefore decrease the composition contrast between the two phases. 
Nevertheless we find that the BIC scores again indicate a regime of strong evidence for phase separation ($0.5 \leq \chi \leq 1$), a regime of weak evidence for two lipid composition populations ($\chi = 0.4, 0.3$), and a regime of complete homogeneity in lipid compositions ($\chi < 0.3$). 
We also observe that within the strong segregation regime the POPC molecules  partition weakly into the liquid-ordered phase, as indicated by the tilted tie-lines in Figure~\ref{fig:phasediag_nochol_nocent_w50}.

\subsection{\label{subsec:LCDsystemsizedependence}Identifying Phase Coexistence via System Size Dependence of Composition Distributions}

\begin{figure}[t]
\includegraphics[width=\columnwidth]{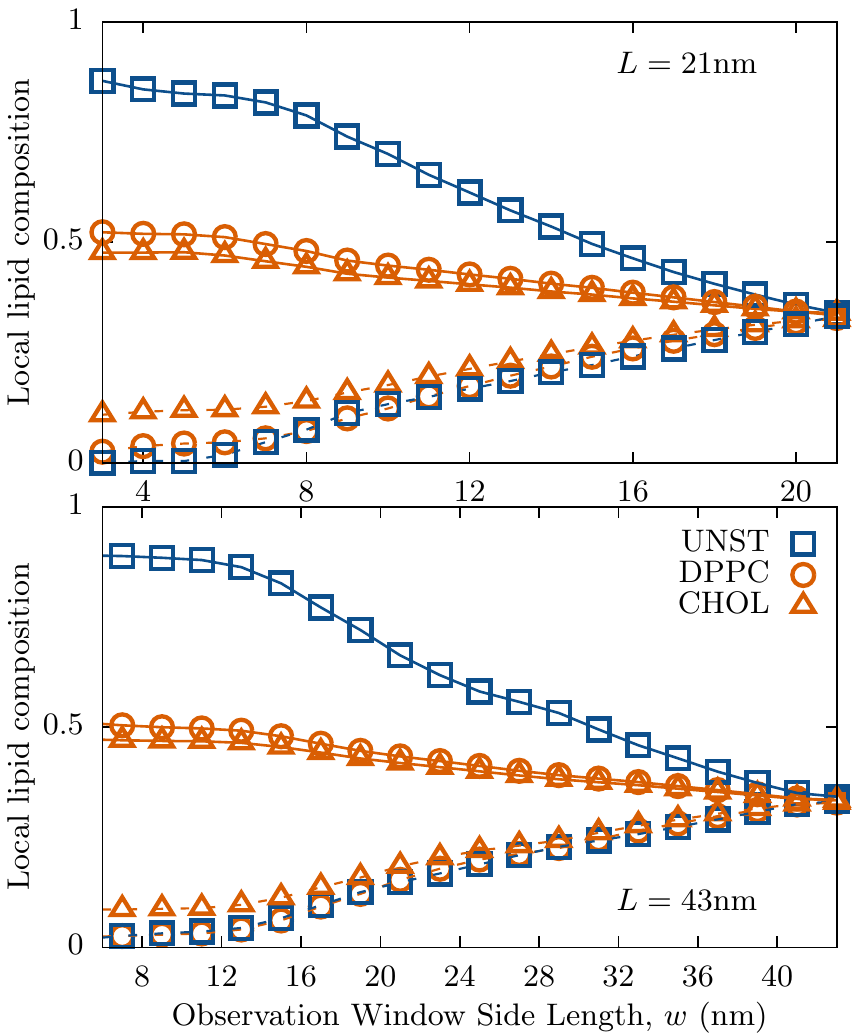}\\
\caption{\label{fig:histpeaksbigsmal} 
Local bilayer compositions as a function of observation window size obtained from the Gaussian Mixture model for a small ($L$ = 21 nm, top) and a large ($L$ = 43 nm, bottom) membrane system at $\chi = 1$. The two graphs are essentially the same, which shows that the length scale of membrane heterogeneity is proportional to the system size, as is expected for thermodynamic phase separation.
}
\end{figure}

In Section~\ref{subsec:LDDobservationsizedependence} we showed that the peak positions in one-dimensional LDDs at large $\chi$ have the dependence on the size $w$ of the observation window that one would expect for a strongly segregated system. To pinpoint the mechanism that causes this segregation we now study the dependence of local distribution functions on the size $L$ of the system. The hallmark characteristic of phase separation is that the associated length scale of heterogeneity is proportional to the system size. In addition, it has recently been shown that phase separation can be artificially suppressed in computer simulations if the size of the system is too small due to the different scaling of mixing entropy and interface energy~\cite{Pantelopulos17}.

To verify that phase separation is indeed the underlying mechanism for the observed heterogeneity in our system we performed an additional set of simulations on much larger membrane systems, obtained by replicating the initial conditions of the previous simulations two-fold in each membrane direction. Once equilibrated, we repeated the Gaussian Mixture analysis described in the previous section on this new dataset.

Figure~\ref{fig:histpeaksbigsmal} compares the compositions obtained from this analysis for both system sizes and a wide range of observation window sizes for a strongly segregated system ($\chi = 1$). The two graphs are essentially the same, even though the ranges of the observation window size are different. At small $w$ we obtain the stable lipid compositions in the two coexisting lipid domains. For this ternary mixture we observe 52\% DPPC, 48\% cholesterol and essentially no DUPC in the liquid-ordered domain, as well as 85\% DUPC, 11\% cholesterol, and 4\% DPPC in the liquid-disordered domain.

As illustrated in the simple continuum model, the apparent peaks in the LDD or LCD functions start to shift toward the bulk density/composition when the sampling window size surpasses the length scale of the heterogeneity. 
For the smaller system we observe that this shift occurs at a length scale of 6 nm, while for the large system it starts when $w$ exceeds 12 nm. This finding indicates that the length scale of heterogeneity of the strongly segregated system grows proportionally with the system size, which is further proof that phase separation is the mechanism that leads to this heterogeneity. 

Finally, we note that the excellent agreement between the two graphs shown in Figure~\ref{fig:histpeaksbigsmal}, which were obtained from independent simulations, indicates that the Gaussian mixture analysis yields robust results for the compositions of coexisting phases.

\section{\label{sec:discussion}Discussion}

The coarse-grained simulations presented in this study capture similar trends along the $\chi$ axis as previous studies on multicomponent lipid bilayers systems in that (1) the scale of lateral heterogeneity, characterized by domain formation, increases with $\chi$, as observed both in experiments \cite{Konyakhina2013, Heberle2013} and simulations \cite{Hakobyan2013, Baoukina2013, Davis2013, Ackerman2015}, (2) the composition difference between coexisting phases also increases with $\chi$, in agreement with previous experimental and computational studies \cite{Konyakhina2013, Ackerman2015}, (3) signatures of phase separation at high $\chi$ such as the overall form of radial distribution functions are in qualitative agreement with previous simulations~\cite{Rosetti2012}, and (4) the phase diagram obtained from our simulations is in qualitative agreement with that obtained in Ref.~\cite{Ackerman2015}.

However, this study also presents new features and perspectives of the problem. In the phase-separated regime, our model-facilitated analysis of the partial radial distribution functions enables us to quantitatively capture the composition information from coexisting lipid phases by relating it to the linear decay of the RDF over long distances. Both the size of the domains and the difference in their compositions can be extracted by comparison with the appropriate continuum model, and the results show a strong dependence of the length scale of bilayer heterogeneity on the lipid composition as described by $\chi$. 
We have shown how the transition from unimodal to bimodal behavior of local distribution functions can be used to identify the onset of phase separation. The system size dependence of these functions is known to contain information about the phase diagram, in particular the precise location of the critical point~\cite{Bruce81, Binder81c}. These functions also depend on the size of the chosen observation window, which we have utilized to verify the onset of phase separation. In addition, Bayesian analysis of Gaussian Mixture models can further enhance the estimates of the phase boundary and of the compositions of the coexisting phases.

For compositions poor in DUPC we have found a well-mixed, homogeneous phase characterized by  short-ranged composition correlations and unimodal distribution functions. This result is at variance with neutron scattering experiments on a similar quaternary lipid mixture, which have revealed domains as small as 10 nanometers in size~\cite{Heberle2013}, and that should therefore be detectable in our simulations. We also find no evidence for a structured fluid phase, such as a microemulsion, in this regime. However, the systems that we simulated are too small to rule out the presence of such a phase. The typical length scale of a bilayer microemulsion is expected to be on the order of 100 nm, which is significantly larger than the systems we can simulate over sufficiently long time scales. The signature of such a phase is a peak in the structure factor at non-zero wavevector. While we do find a significant increase in $S(k)$ at small $k$, our observations do not reach small enough wavevectors to determine whether this increase continues monotonically or whether it reverses, which would yield the characteristic peak.

Further studies, both experimental and computational, will be required to fully elucidate the nature of this material, which at least on the small length scales studied in our simulations seems homogeneous. The results presented in this work contribute to this quest by using novel approaches to accurately determine the boundaries of the phase coexistence region in this quaternary lipid bilayer.

\section*{Acknowledgments}

This work was facilitated though the use of advanced computational, storage, and networking infrastructure provided by the Hyak supercomputer system at the University of Washington.

\bibliography{paper}

\begin{thebibliography}{68}%
\makeatletter
\providecommand \@ifxundefined [1]{%
 \@ifx{#1\undefined}
}%
\providecommand \@ifnum [1]{%
 \ifnum #1\expandafter \@firstoftwo
 \else \expandafter \@secondoftwo
 \fi
}%
\providecommand \@ifx [1]{%
 \ifx #1\expandafter \@firstoftwo
 \else \expandafter \@secondoftwo
 \fi
}%
\providecommand \natexlab [1]{#1}%
\providecommand \enquote  [1]{``#1''}%
\providecommand \bibnamefont  [1]{#1}%
\providecommand \bibfnamefont [1]{#1}%
\providecommand \citenamefont [1]{#1}%
\providecommand \href@noop [0]{\@secondoftwo}%
\providecommand \href [0]{\begingroup \@sanitize@url \@href}%
\providecommand \@href[1]{\@@startlink{#1}\@@href}%
\providecommand \@@href[1]{\endgroup#1\@@endlink}%
\providecommand \@sanitize@url [0]{\catcode `\\12\catcode `\$12\catcode
  `\&12\catcode `\#12\catcode `\^12\catcode `\_12\catcode `\%12\relax}%
\providecommand \@@startlink[1]{}%
\providecommand \@@endlink[0]{}%
\providecommand \url  [0]{\begingroup\@sanitize@url \@url }%
\providecommand \@url [1]{\endgroup\@href {#1}{\urlprefix }}%
\providecommand \urlprefix  [0]{URL }%
\providecommand \Eprint [0]{\href }%
\providecommand \doibase [0]{http://dx.doi.org/}%
\providecommand \selectlanguage [0]{\@gobble}%
\providecommand \bibinfo  [0]{\@secondoftwo}%
\providecommand \bibfield  [0]{\@secondoftwo}%
\providecommand \translation [1]{[#1]}%
\providecommand \BibitemOpen [0]{}%
\providecommand \bibitemStop [0]{}%
\providecommand \bibitemNoStop [0]{.\EOS\space}%
\providecommand \EOS [0]{\spacefactor3000\relax}%
\providecommand \BibitemShut  [1]{\csname bibitem#1\endcsname}%
\let\auto@bib@innerbib\@empty
\bibitem [{\citenamefont {Simons}\ and\ \citenamefont
  {Ikonen}(1997)}]{Simons1997}%
  \BibitemOpen
  \bibfield  {author} {\bibinfo {author} {\bibfnamefont {K.}~\bibnamefont
  {Simons}}\ and\ \bibinfo {author} {\bibfnamefont {E.}~\bibnamefont
  {Ikonen}},\ }\href {\doibase 10.1038/42408} {\bibfield  {journal} {\bibinfo
  {journal} {Nature}\ }\textbf {\bibinfo {volume} {387}},\ \bibinfo {pages}
  {569} (\bibinfo {year} {1997})}\BibitemShut {NoStop}%
\bibitem [{\citenamefont {Simons}\ and\ \citenamefont
  {Vaz}(2004)}]{Simons2004}%
  \BibitemOpen
  \bibfield  {author} {\bibinfo {author} {\bibfnamefont {K.}~\bibnamefont
  {Simons}}\ and\ \bibinfo {author} {\bibfnamefont {W.~L.~C.}\ \bibnamefont
  {Vaz}},\ }\href {\doibase 10.1146/annurev.biophys.32.110601.141803}
  {\bibfield  {journal} {\bibinfo  {journal} {Annual review of biophysics and
  biomolecular structure}\ }\textbf {\bibinfo {volume} {33}},\ \bibinfo {pages}
  {269} (\bibinfo {year} {2004})}\BibitemShut {NoStop}%
\bibitem [{\citenamefont {Simons}\ and\ \citenamefont
  {Sampaio}(2011)}]{Simons2011}%
  \BibitemOpen
  \bibfield  {author} {\bibinfo {author} {\bibfnamefont {K.}~\bibnamefont
  {Simons}}\ and\ \bibinfo {author} {\bibfnamefont {J.~L.}\ \bibnamefont
  {Sampaio}},\ }\href {\doibase 10.1101/cshperspect.a004697} {\bibfield
  {journal} {\bibinfo  {journal} {Cold Spring Harbor perspectives in biology}\
  }\textbf {\bibinfo {volume} {3:a004697}},\ \bibinfo {pages} {1} (\bibinfo
  {year} {2011})}\BibitemShut {NoStop}%
\bibitem [{\citenamefont {Lingwood}\ and\ \citenamefont
  {Simons}(2010)}]{Lingwood2010}%
  \BibitemOpen
  \bibfield  {author} {\bibinfo {author} {\bibfnamefont {D.}~\bibnamefont
  {Lingwood}}\ and\ \bibinfo {author} {\bibfnamefont {K.}~\bibnamefont
  {Simons}},\ }\href {\doibase 10.1126/science.1174621} {\bibfield  {journal}
  {\bibinfo  {journal} {Science (New York, N.Y.)}\ }\textbf {\bibinfo {volume}
  {327}},\ \bibinfo {pages} {46} (\bibinfo {year} {2010})}\BibitemShut
  {NoStop}%
\bibitem [{\citenamefont {Anderson}(2002)}]{Anderson2002}%
  \BibitemOpen
  \bibfield  {author} {\bibinfo {author} {\bibfnamefont {R.~G.~W.}\
  \bibnamefont {Anderson}},\ }\href {\doibase 10.1126/science.1068886}
  {\bibfield  {journal} {\bibinfo  {journal} {Science}\ }\textbf {\bibinfo
  {volume} {296}},\ \bibinfo {pages} {1821} (\bibinfo {year}
  {2002})}\BibitemShut {NoStop}%
\bibitem [{\citenamefont {Toulmay}\ and\ \citenamefont
  {Prinz}(2012)}]{Toulmay2012}%
  \BibitemOpen
  \bibfield  {author} {\bibinfo {author} {\bibfnamefont {A.}~\bibnamefont
  {Toulmay}}\ and\ \bibinfo {author} {\bibfnamefont {W.~a.}\ \bibnamefont
  {Prinz}},\ }\href {\doibase 10.1242/jcs.085118} {\bibfield  {journal}
  {\bibinfo  {journal} {Journal of cell science}\ }\textbf {\bibinfo {volume}
  {125}},\ \bibinfo {pages} {49} (\bibinfo {year} {2012})}\BibitemShut
  {NoStop}%
\bibitem [{\citenamefont {Diaz-Rohrer}\ \emph
  {et~al.}(2014{\natexlab{a}})\citenamefont {Diaz-Rohrer}, \citenamefont
  {Levental},\ and\ \citenamefont {Levental}}]{Diaz-Rohrer2014a}%
  \BibitemOpen
  \bibfield  {author} {\bibinfo {author} {\bibfnamefont {B.~B.}\ \bibnamefont
  {Diaz-Rohrer}}, \bibinfo {author} {\bibfnamefont {K.~R.}\ \bibnamefont
  {Levental}}, \ and\ \bibinfo {author} {\bibfnamefont {I.}~\bibnamefont
  {Levental}},\ }\href {\doibase 10.1016/j.bbamem.2014.07.029} {\bibfield
  {journal} {\bibinfo  {journal} {Biochimica et Biophysica Acta -
  Biomembranes}\ }\textbf {\bibinfo {volume} {1838}},\ \bibinfo {pages} {3003}
  (\bibinfo {year} {2014}{\natexlab{a}})}\BibitemShut {NoStop}%
\bibitem [{\citenamefont {Levental}\ and\ \citenamefont
  {Veatch}(2016)}]{Levental2016}%
  \BibitemOpen
  \bibfield  {author} {\bibinfo {author} {\bibfnamefont {I.}~\bibnamefont
  {Levental}}\ and\ \bibinfo {author} {\bibfnamefont {S.~L.}\ \bibnamefont
  {Veatch}},\ }\href {\doibase 10.1016/j.jmb.2016.08.022} {\bibfield  {journal}
  {\bibinfo  {journal} {Journal of Molecular Biology}\ }\textbf {\bibinfo
  {volume} {428}},\ \bibinfo {pages} {4749} (\bibinfo {year}
  {2016})}\BibitemShut {NoStop}%
\bibitem [{\citenamefont {Rajendran}(2005)}]{Rajendran2005}%
  \BibitemOpen
  \bibfield  {author} {\bibinfo {author} {\bibfnamefont {L.}~\bibnamefont
  {Rajendran}},\ }\href {\doibase 10.1242/jcs.01681} {\bibfield  {journal}
  {\bibinfo  {journal} {Journal of Cell Science}\ }\textbf {\bibinfo {volume}
  {118}},\ \bibinfo {pages} {1099} (\bibinfo {year} {2005})}\BibitemShut
  {NoStop}%
\bibitem [{\citenamefont {Simons}\ and\ \citenamefont
  {Gerl}(2010)}]{Simons2010}%
  \BibitemOpen
  \bibfield  {author} {\bibinfo {author} {\bibfnamefont {K.}~\bibnamefont
  {Simons}}\ and\ \bibinfo {author} {\bibfnamefont {M.~J.}\ \bibnamefont
  {Gerl}},\ }\href {\doibase 10.1038/nrm2977} {\bibfield  {journal} {\bibinfo
  {journal} {Nature reviews. Molecular cell biology}\ }\textbf {\bibinfo
  {volume} {11}},\ \bibinfo {pages} {688} (\bibinfo {year} {2010})}\BibitemShut
  {NoStop}%
\bibitem [{\citenamefont {Veatch}\ and\ \citenamefont
  {Keller}(2005)}]{Veatch2005}%
  \BibitemOpen
  \bibfield  {author} {\bibinfo {author} {\bibfnamefont {S.~L.}\ \bibnamefont
  {Veatch}}\ and\ \bibinfo {author} {\bibfnamefont {S.~L.}\ \bibnamefont
  {Keller}},\ }\href {\doibase 10.1016/j.bbamcr.2005.06.010} {\bibfield
  {journal} {\bibinfo  {journal} {Biochimica et Biophysica Acta - Molecular
  Cell Research}\ }\textbf {\bibinfo {volume} {1746}},\ \bibinfo {pages} {172}
  (\bibinfo {year} {2005})}\BibitemShut {NoStop}%
\bibitem [{\citenamefont {Ing{\'{o}}lfsson}\ \emph {et~al.}(2014)\citenamefont
  {Ing{\'{o}}lfsson}, \citenamefont {Melo}, \citenamefont {{Van Eerden}},
  \citenamefont {Arnarez}, \citenamefont {Lopez}, \citenamefont {Wassenaar},
  \citenamefont {Periole}, \citenamefont {{De Vries}}, \citenamefont
  {Tieleman},\ and\ \citenamefont {Marrink}}]{Ingolfsson2014}%
  \BibitemOpen
  \bibfield  {author} {\bibinfo {author} {\bibfnamefont {H.~I.}\ \bibnamefont
  {Ing{\'{o}}lfsson}}, \bibinfo {author} {\bibfnamefont {M.~N.}\ \bibnamefont
  {Melo}}, \bibinfo {author} {\bibfnamefont {F.~J.}\ \bibnamefont {{Van
  Eerden}}}, \bibinfo {author} {\bibfnamefont {C.}~\bibnamefont {Arnarez}},
  \bibinfo {author} {\bibfnamefont {C.~A.}\ \bibnamefont {Lopez}}, \bibinfo
  {author} {\bibfnamefont {T.~A.}\ \bibnamefont {Wassenaar}}, \bibinfo {author}
  {\bibfnamefont {X.}~\bibnamefont {Periole}}, \bibinfo {author} {\bibfnamefont
  {A.~H.}\ \bibnamefont {{De Vries}}}, \bibinfo {author} {\bibfnamefont
  {D.~P.}\ \bibnamefont {Tieleman}}, \ and\ \bibinfo {author} {\bibfnamefont
  {S.~J.}\ \bibnamefont {Marrink}},\ }\href {\doibase 10.1021/ja507832e}
  {\bibfield  {journal} {\bibinfo  {journal} {Journal of the American Chemical
  Society}\ }\textbf {\bibinfo {volume} {136}},\ \bibinfo {pages} {14554}
  (\bibinfo {year} {2014})}\BibitemShut {NoStop}%
\bibitem [{\citenamefont {Veatch}\ and\ \citenamefont
  {Keller}(2002)}]{Veatch2002}%
  \BibitemOpen
  \bibfield  {author} {\bibinfo {author} {\bibfnamefont {S.~L.}\ \bibnamefont
  {Veatch}}\ and\ \bibinfo {author} {\bibfnamefont {S.~L.}\ \bibnamefont
  {Keller}},\ }\href {\doibase 10.1103/PhysRevLett.89.268101} {\bibfield
  {journal} {\bibinfo  {journal} {Physical Review Letters}\ }\textbf {\bibinfo
  {volume} {89}},\ \bibinfo {pages} {268101} (\bibinfo {year}
  {2002})}\BibitemShut {NoStop}%
\bibitem [{\citenamefont {Veatch}\ and\ \citenamefont
  {Keller}(2003)}]{Veatch2003}%
  \BibitemOpen
  \bibfield  {author} {\bibinfo {author} {\bibfnamefont {S.~L.}\ \bibnamefont
  {Veatch}}\ and\ \bibinfo {author} {\bibfnamefont {S.~L.}\ \bibnamefont
  {Keller}},\ }\href {\doibase 10.1016/S0006-3495(03)74726-2} {\bibfield
  {journal} {\bibinfo  {journal} {Biophysical journal}\ }\textbf {\bibinfo
  {volume} {85}},\ \bibinfo {pages} {3074} (\bibinfo {year}
  {2003})}\BibitemShut {NoStop}%
\bibitem [{\citenamefont {Zhao}\ \emph {et~al.}(2007)\citenamefont {Zhao},
  \citenamefont {Wu}, \citenamefont {Heberle}, \citenamefont {Mills},
  \citenamefont {Klawitter}, \citenamefont {Huang}, \citenamefont {Costanza},\
  and\ \citenamefont {Feigenson}}]{Zhao2007}%
  \BibitemOpen
  \bibfield  {author} {\bibinfo {author} {\bibfnamefont {J.}~\bibnamefont
  {Zhao}}, \bibinfo {author} {\bibfnamefont {J.}~\bibnamefont {Wu}}, \bibinfo
  {author} {\bibfnamefont {F.~A.}\ \bibnamefont {Heberle}}, \bibinfo {author}
  {\bibfnamefont {T.~T.}\ \bibnamefont {Mills}}, \bibinfo {author}
  {\bibfnamefont {P.}~\bibnamefont {Klawitter}}, \bibinfo {author}
  {\bibfnamefont {G.}~\bibnamefont {Huang}}, \bibinfo {author} {\bibfnamefont
  {G.}~\bibnamefont {Costanza}}, \ and\ \bibinfo {author} {\bibfnamefont
  {G.~W.}\ \bibnamefont {Feigenson}},\ }\href {\doibase
  10.1016/j.bbamem.2007.07.008} {\bibfield  {journal} {\bibinfo  {journal}
  {Biochimica et Biophysica Acta - Biomembranes}\ }\textbf {\bibinfo {volume}
  {1768}},\ \bibinfo {pages} {2764} (\bibinfo {year} {2007})}\BibitemShut
  {NoStop}%
\bibitem [{\citenamefont {Konyakhina}\ \emph {et~al.}(2011)\citenamefont
  {Konyakhina}, \citenamefont {Goh}, \citenamefont {Amazon}, \citenamefont
  {Heberle}, \citenamefont {Wu},\ and\ \citenamefont
  {Feigenson}}]{Konyakhina2011}%
  \BibitemOpen
  \bibfield  {author} {\bibinfo {author} {\bibfnamefont {T.~M.}\ \bibnamefont
  {Konyakhina}}, \bibinfo {author} {\bibfnamefont {S.~L.}\ \bibnamefont {Goh}},
  \bibinfo {author} {\bibfnamefont {J.}~\bibnamefont {Amazon}}, \bibinfo
  {author} {\bibfnamefont {F.~A.}\ \bibnamefont {Heberle}}, \bibinfo {author}
  {\bibfnamefont {J.}~\bibnamefont {Wu}}, \ and\ \bibinfo {author}
  {\bibfnamefont {G.~W.}\ \bibnamefont {Feigenson}},\ }\href {\doibase
  10.1016/j.bpj.2011.06.019} {\bibfield  {journal} {\bibinfo  {journal}
  {Biophysical Journal}\ }\textbf {\bibinfo {volume} {101}},\ \bibinfo {pages}
  {L8} (\bibinfo {year} {2011})}\BibitemShut {NoStop}%
\bibitem [{\citenamefont {Toulmay}\ and\ \citenamefont
  {Prinz}(2013)}]{Toulmay2013}%
  \BibitemOpen
  \bibfield  {author} {\bibinfo {author} {\bibfnamefont {A.}~\bibnamefont
  {Toulmay}}\ and\ \bibinfo {author} {\bibfnamefont {W.~A.}\ \bibnamefont
  {Prinz}},\ }\href {\doibase 10.1083/jcb.201301039} {\bibfield  {journal}
  {\bibinfo  {journal} {Journal of Cell Biology}\ }\textbf {\bibinfo {volume}
  {202}},\ \bibinfo {pages} {35} (\bibinfo {year} {2013})}\BibitemShut
  {NoStop}%
\bibitem [{\citenamefont {Diaz-Rohrer}\ \emph
  {et~al.}(2014{\natexlab{b}})\citenamefont {Diaz-Rohrer}, \citenamefont
  {Levental}, \citenamefont {Simons},\ and\ \citenamefont
  {Levental}}]{Diaz-Rohrer2014}%
  \BibitemOpen
  \bibfield  {author} {\bibinfo {author} {\bibfnamefont {B.~B.}\ \bibnamefont
  {Diaz-Rohrer}}, \bibinfo {author} {\bibfnamefont {K.~R.}\ \bibnamefont
  {Levental}}, \bibinfo {author} {\bibfnamefont {K.}~\bibnamefont {Simons}}, \
  and\ \bibinfo {author} {\bibfnamefont {I.}~\bibnamefont {Levental}},\ }\href
  {\doibase 10.1073/pnas.1404582111} {\bibfield  {journal} {\bibinfo  {journal}
  {Proceedings of the National Academy of Sciences of the United States of
  America}\ }\textbf {\bibinfo {volume} {111}},\ \bibinfo {pages} {8500}
  (\bibinfo {year} {2014}{\natexlab{b}})}\BibitemShut {NoStop}%
\bibitem [{\citenamefont {Holowka}\ \emph {et~al.}(2005)\citenamefont
  {Holowka}, \citenamefont {Gosse}, \citenamefont {Hammond}, \citenamefont
  {Han}, \citenamefont {Sengupta}, \citenamefont {Smith}, \citenamefont
  {Wagenknecht-Wiesner}, \citenamefont {Wu}, \citenamefont {Young},\ and\
  \citenamefont {Baird}}]{Holowka2005}%
  \BibitemOpen
  \bibfield  {author} {\bibinfo {author} {\bibfnamefont {D.}~\bibnamefont
  {Holowka}}, \bibinfo {author} {\bibfnamefont {J.~A.}\ \bibnamefont {Gosse}},
  \bibinfo {author} {\bibfnamefont {A.~T.}\ \bibnamefont {Hammond}}, \bibinfo
  {author} {\bibfnamefont {X.}~\bibnamefont {Han}}, \bibinfo {author}
  {\bibfnamefont {P.}~\bibnamefont {Sengupta}}, \bibinfo {author}
  {\bibfnamefont {N.~L.}\ \bibnamefont {Smith}}, \bibinfo {author}
  {\bibfnamefont {A.}~\bibnamefont {Wagenknecht-Wiesner}}, \bibinfo {author}
  {\bibfnamefont {M.}~\bibnamefont {Wu}}, \bibinfo {author} {\bibfnamefont
  {R.~M.}\ \bibnamefont {Young}}, \ and\ \bibinfo {author} {\bibfnamefont
  {B.}~\bibnamefont {Baird}},\ }\href {\doibase 10.1016/j.bbamcr.2005.06.007}
  {\bibfield  {journal} {\bibinfo  {journal} {Biochimica et Biophysica Acta -
  Molecular Cell Research}\ }\textbf {\bibinfo {volume} {1746}},\ \bibinfo
  {pages} {252} (\bibinfo {year} {2005})}\BibitemShut {NoStop}%
\bibitem [{\citenamefont {Sengupta}\ \emph
  {et~al.}(2007{\natexlab{a}})\citenamefont {Sengupta}, \citenamefont {Baird},\
  and\ \citenamefont {Holowka}}]{Sengupta2007}%
  \BibitemOpen
  \bibfield  {author} {\bibinfo {author} {\bibfnamefont {P.}~\bibnamefont
  {Sengupta}}, \bibinfo {author} {\bibfnamefont {B.}~\bibnamefont {Baird}}, \
  and\ \bibinfo {author} {\bibfnamefont {D.}~\bibnamefont {Holowka}},\ }\href
  {\doibase S1084-9521(07)00100-0 [pii]\r10.1016/j.semcdb.2007.07.010}
  {\bibfield  {journal} {\bibinfo  {journal} {Semin Cell Dev Biol}\ }\textbf
  {\bibinfo {volume} {18}},\ \bibinfo {pages} {583} (\bibinfo {year}
  {2007}{\natexlab{a}})}\BibitemShut {NoStop}%
\bibitem [{\citenamefont {Baumgart}\ \emph {et~al.}(2003)\citenamefont
  {Baumgart}, \citenamefont {Hess},\ and\ \citenamefont {Webb}}]{Baumgart2003}%
  \BibitemOpen
  \bibfield  {author} {\bibinfo {author} {\bibfnamefont {T.}~\bibnamefont
  {Baumgart}}, \bibinfo {author} {\bibfnamefont {S.~T.}\ \bibnamefont {Hess}},
  \ and\ \bibinfo {author} {\bibfnamefont {W.~W.}\ \bibnamefont {Webb}},\
  }\href {\doibase 10.1038/nature02013} {\bibfield  {journal} {\bibinfo
  {journal} {Nature}\ }\textbf {\bibinfo {volume} {425}},\ \bibinfo {pages}
  {821} (\bibinfo {year} {2003})}\BibitemShut {NoStop}%
\bibitem [{\citenamefont {Marsh}(2009)}]{Marsh2009}%
  \BibitemOpen
  \bibfield  {author} {\bibinfo {author} {\bibfnamefont {D.}~\bibnamefont
  {Marsh}},\ }\href {\doibase 10.1016/j.bbamem.2009.08.004} {\bibfield
  {journal} {\bibinfo  {journal} {Biochim. Biophys. Acta -- Biomembranes}\
  }\textbf {\bibinfo {volume} {1788}},\ \bibinfo {pages} {2114} (\bibinfo
  {year} {2009})}\BibitemShut {NoStop}%
\bibitem [{\citenamefont {Heberle}\ \emph {et~al.}(2010)\citenamefont
  {Heberle}, \citenamefont {Wu}, \citenamefont {Goh}, \citenamefont
  {Petruzielo},\ and\ \citenamefont {Feigenson}}]{Heberle2010}%
  \BibitemOpen
  \bibfield  {author} {\bibinfo {author} {\bibfnamefont {F.~A.}\ \bibnamefont
  {Heberle}}, \bibinfo {author} {\bibfnamefont {J.}~\bibnamefont {Wu}},
  \bibinfo {author} {\bibfnamefont {S.~L.}\ \bibnamefont {Goh}}, \bibinfo
  {author} {\bibfnamefont {R.~S.}\ \bibnamefont {Petruzielo}}, \ and\ \bibinfo
  {author} {\bibfnamefont {G.~W.}\ \bibnamefont {Feigenson}},\ }\href {\doibase
  10.1016/j.bpj.2010.09.064} {\bibfield  {journal} {\bibinfo  {journal}
  {Biophysical Journal}\ }\textbf {\bibinfo {volume} {99}},\ \bibinfo {pages}
  {3309} (\bibinfo {year} {2010})}\BibitemShut {NoStop}%
\bibitem [{\citenamefont {Baumgart}\ \emph {et~al.}(2007)\citenamefont
  {Baumgart}, \citenamefont {Hammond}, \citenamefont {Sengupta}, \citenamefont
  {Hess}, \citenamefont {Holowka}, \citenamefont {Baird},\ and\ \citenamefont
  {Webb}}]{Baumgart2007}%
  \BibitemOpen
  \bibfield  {author} {\bibinfo {author} {\bibfnamefont {T.}~\bibnamefont
  {Baumgart}}, \bibinfo {author} {\bibfnamefont {A.~T.}\ \bibnamefont
  {Hammond}}, \bibinfo {author} {\bibfnamefont {P.}~\bibnamefont {Sengupta}},
  \bibinfo {author} {\bibfnamefont {S.~T.}\ \bibnamefont {Hess}}, \bibinfo
  {author} {\bibfnamefont {D.~A.}\ \bibnamefont {Holowka}}, \bibinfo {author}
  {\bibfnamefont {B.~A.}\ \bibnamefont {Baird}}, \ and\ \bibinfo {author}
  {\bibfnamefont {W.~W.}\ \bibnamefont {Webb}},\ }\href@noop {} {\ \textbf
  {\bibinfo {volume} {104}} (\bibinfo {year} {2007})}\BibitemShut {NoStop}%
\bibitem [{\citenamefont {Veatch}\ \emph {et~al.}(2008)\citenamefont {Veatch},
  \citenamefont {Cicuta}, \citenamefont {Sengupta}, \citenamefont
  {Honerkamp-Smith}, \citenamefont {Holowka},\ and\ \citenamefont
  {Baird}}]{Veatch2008}%
  \BibitemOpen
  \bibfield  {author} {\bibinfo {author} {\bibfnamefont {S.~L.}\ \bibnamefont
  {Veatch}}, \bibinfo {author} {\bibfnamefont {P.}~\bibnamefont {Cicuta}},
  \bibinfo {author} {\bibfnamefont {P.}~\bibnamefont {Sengupta}}, \bibinfo
  {author} {\bibfnamefont {A.}~\bibnamefont {Honerkamp-Smith}}, \bibinfo
  {author} {\bibfnamefont {D.}~\bibnamefont {Holowka}}, \ and\ \bibinfo
  {author} {\bibfnamefont {B.}~\bibnamefont {Baird}},\ }\href {\doibase
  10.1021/cb800012x} {\bibfield  {journal} {\bibinfo  {journal} {ACS Chemical
  Biology}\ }\textbf {\bibinfo {volume} {3}},\ \bibinfo {pages} {287} (\bibinfo
  {year} {2008})}\BibitemShut {NoStop}%
\bibitem [{\citenamefont {Sengupta}\ \emph
  {et~al.}(2007{\natexlab{b}})\citenamefont {Sengupta}, \citenamefont
  {Holowka},\ and\ \citenamefont {Baird}}]{Sengupta2007a}%
  \BibitemOpen
  \bibfield  {author} {\bibinfo {author} {\bibfnamefont {P.}~\bibnamefont
  {Sengupta}}, \bibinfo {author} {\bibfnamefont {D.}~\bibnamefont {Holowka}}, \
  and\ \bibinfo {author} {\bibfnamefont {B.}~\bibnamefont {Baird}},\ }\href
  {\doibase 10.1529/biophysj.106.094730} {\bibfield  {journal} {\bibinfo
  {journal} {Biophysical Journal}\ }\textbf {\bibinfo {volume} {92}},\ \bibinfo
  {pages} {3564} (\bibinfo {year} {2007}{\natexlab{b}})}\BibitemShut {NoStop}%
\bibitem [{\citenamefont {Edidin}(2003)}]{Edidin2003}%
  \BibitemOpen
  \bibfield  {author} {\bibinfo {author} {\bibfnamefont {M.}~\bibnamefont
  {Edidin}},\ }\href {\doibase 10.1146/annurev.biophys.32.110601.142439}
  {\bibfield  {journal} {\bibinfo  {journal} {Annual review of biophysics and
  biomolecular structure}\ }\textbf {\bibinfo {volume} {32}},\ \bibinfo {pages}
  {257} (\bibinfo {year} {2003})}\BibitemShut {NoStop}%
\bibitem [{\citenamefont {Stone}\ \emph {et~al.}(2017)\citenamefont {Stone},
  \citenamefont {Shelby},\ and\ \citenamefont {Veatch}}]{Stone2017}%
  \BibitemOpen
  \bibfield  {author} {\bibinfo {author} {\bibfnamefont {M.~B.}\ \bibnamefont
  {Stone}}, \bibinfo {author} {\bibfnamefont {S.~A.}\ \bibnamefont {Shelby}}, \
  and\ \bibinfo {author} {\bibfnamefont {S.~L.}\ \bibnamefont {Veatch}},\
  }\href {\doibase 10.1021/acs.chemrev.6b00716} {\bibfield  {journal} {\bibinfo
   {journal} {Chemical Reviews}\ }\textbf {\bibinfo {volume} {117}},\ \bibinfo
  {pages} {7457} (\bibinfo {year} {2017})}\BibitemShut {NoStop}%
\bibitem [{\citenamefont {Petruzielo}\ \emph {et~al.}(2013)\citenamefont
  {Petruzielo}, \citenamefont {Heberle}, \citenamefont {Drazba}, \citenamefont
  {Katsaras},\ and\ \citenamefont {Feigenson}}]{Petruzielo2013}%
  \BibitemOpen
  \bibfield  {author} {\bibinfo {author} {\bibfnamefont {R.~S.}\ \bibnamefont
  {Petruzielo}}, \bibinfo {author} {\bibfnamefont {F.~A.}\ \bibnamefont
  {Heberle}}, \bibinfo {author} {\bibfnamefont {P.}~\bibnamefont {Drazba}},
  \bibinfo {author} {\bibfnamefont {J.}~\bibnamefont {Katsaras}}, \ and\
  \bibinfo {author} {\bibfnamefont {G.~W.}\ \bibnamefont {Feigenson}},\ }\href
  {\doibase 10.1016/j.bbamem.2013.01.007} {\bibfield  {journal} {\bibinfo
  {journal} {Biochimica et Biophysica Acta - Biomembranes}\ }\textbf {\bibinfo
  {volume} {1828}},\ \bibinfo {pages} {1302} (\bibinfo {year}
  {2013})}\BibitemShut {NoStop}%
\bibitem [{\citenamefont {Konyakhina}\ \emph {et~al.}(2013)\citenamefont
  {Konyakhina}, \citenamefont {Wu}, \citenamefont {Mastroianni}, \citenamefont
  {Heberle},\ and\ \citenamefont {Feigenson}}]{Konyakhina2013}%
  \BibitemOpen
  \bibfield  {author} {\bibinfo {author} {\bibfnamefont {T.~M.}\ \bibnamefont
  {Konyakhina}}, \bibinfo {author} {\bibfnamefont {J.}~\bibnamefont {Wu}},
  \bibinfo {author} {\bibfnamefont {J.~D.}\ \bibnamefont {Mastroianni}},
  \bibinfo {author} {\bibfnamefont {F.~A.}\ \bibnamefont {Heberle}}, \ and\
  \bibinfo {author} {\bibfnamefont {G.~W.}\ \bibnamefont {Feigenson}},\ }\href
  {\doibase 10.1016/j.bbamem.2013.05.020} {\bibfield  {journal} {\bibinfo
  {journal} {Biochimica et Biophysica Acta - Biomembranes}\ }\textbf {\bibinfo
  {volume} {1828}},\ \bibinfo {pages} {2204} (\bibinfo {year}
  {2013})}\BibitemShut {NoStop}%
\bibitem [{\citenamefont {Heberle}\ \emph
  {et~al.}(2013{\natexlab{a}})\citenamefont {Heberle}, \citenamefont
  {Doktorova}, \citenamefont {Goh}, \citenamefont {Standaert}, \citenamefont
  {Katsaras},\ and\ \citenamefont {Feigenson}}]{Heberle2013a}%
  \BibitemOpen
  \bibfield  {author} {\bibinfo {author} {\bibfnamefont {F.~A.}\ \bibnamefont
  {Heberle}}, \bibinfo {author} {\bibfnamefont {M.}~\bibnamefont {Doktorova}},
  \bibinfo {author} {\bibfnamefont {S.~L.}\ \bibnamefont {Goh}}, \bibinfo
  {author} {\bibfnamefont {R.~F.}\ \bibnamefont {Standaert}}, \bibinfo {author}
  {\bibfnamefont {J.}~\bibnamefont {Katsaras}}, \ and\ \bibinfo {author}
  {\bibfnamefont {G.~W.}\ \bibnamefont {Feigenson}},\ }\href {\doibase
  10.1021/ja407624c} {\bibfield  {journal} {\bibinfo  {journal} {Journal of the
  American Chemical Society}\ }\textbf {\bibinfo {volume} {135}},\ \bibinfo
  {pages} {14932} (\bibinfo {year} {2013}{\natexlab{a}})}\BibitemShut {NoStop}%
\bibitem [{\citenamefont {Lindblom}\ and\ \citenamefont
  {Or{\"{a}}dd}(2009)}]{Lindblom2009}%
  \BibitemOpen
  \bibfield  {author} {\bibinfo {author} {\bibfnamefont {G.}~\bibnamefont
  {Lindblom}}\ and\ \bibinfo {author} {\bibfnamefont {G.}~\bibnamefont
  {Or{\"{a}}dd}},\ }\href {\doibase 10.1016/j.bbamem.2008.08.016} {\bibfield
  {journal} {\bibinfo  {journal} {Biochimica et Biophysica Acta -
  Biomembranes}\ }\textbf {\bibinfo {volume} {1788}},\ \bibinfo {pages} {234}
  (\bibinfo {year} {2009})}\BibitemShut {NoStop}%
\bibitem [{\citenamefont {Leftin}\ \emph {et~al.}(2014)\citenamefont {Leftin},
  \citenamefont {Molugu}, \citenamefont {Job}, \citenamefont {Beyer},\ and\
  \citenamefont {Brown}}]{Leftin2014}%
  \BibitemOpen
  \bibfield  {author} {\bibinfo {author} {\bibfnamefont {A.}~\bibnamefont
  {Leftin}}, \bibinfo {author} {\bibfnamefont {T.~R.}\ \bibnamefont {Molugu}},
  \bibinfo {author} {\bibfnamefont {C.}~\bibnamefont {Job}}, \bibinfo {author}
  {\bibfnamefont {K.}~\bibnamefont {Beyer}}, \ and\ \bibinfo {author}
  {\bibfnamefont {M.~F.}\ \bibnamefont {Brown}},\ }\href {\doibase
  10.1016/j.bpj.2014.07.044} {\bibfield  {journal} {\bibinfo  {journal}
  {Biophysical Journal}\ }\textbf {\bibinfo {volume} {107}},\ \bibinfo {pages}
  {2274} (\bibinfo {year} {2014})}\BibitemShut {NoStop}%
\bibitem [{\citenamefont {Gr{\'{e}}lard}\ \emph {et~al.}(2010)\citenamefont
  {Gr{\'{e}}lard}, \citenamefont {Loudet}, \citenamefont {Diller},\ and\
  \citenamefont {Dufourc}}]{Grelard2010}%
  \BibitemOpen
  \bibfield  {author} {\bibinfo {author} {\bibfnamefont {A.}~\bibnamefont
  {Gr{\'{e}}lard}}, \bibinfo {author} {\bibfnamefont {C.}~\bibnamefont
  {Loudet}}, \bibinfo {author} {\bibfnamefont {A.}~\bibnamefont {Diller}}, \
  and\ \bibinfo {author} {\bibfnamefont {E.~J.}\ \bibnamefont {Dufourc}}\
  }(\bibinfo {year} {2010})\ Chap.\ \bibinfo {chapter} {18 NMR Spe}, pp.\
  \bibinfo {pages} {341--359}\BibitemShut {NoStop}%
\bibitem [{\citenamefont {Ando}\ \emph {et~al.}(2015)\citenamefont {Ando},
  \citenamefont {Kinoshita}, \citenamefont {Cui}, \citenamefont {Yamakoshi},
  \citenamefont {Dodo},\ and\ \citenamefont {Fujita}}]{Ando2015}%
  \BibitemOpen
  \bibfield  {author} {\bibinfo {author} {\bibfnamefont {J.}~\bibnamefont
  {Ando}}, \bibinfo {author} {\bibfnamefont {M.}~\bibnamefont {Kinoshita}},
  \bibinfo {author} {\bibfnamefont {J.}~\bibnamefont {Cui}}, \bibinfo {author}
  {\bibfnamefont {H.}~\bibnamefont {Yamakoshi}}, \bibinfo {author}
  {\bibfnamefont {K.}~\bibnamefont {Dodo}}, \ and\ \bibinfo {author}
  {\bibfnamefont {K.}~\bibnamefont {Fujita}},\ }\href {\doibase
  10.1073/pnas.1418088112} {\bibfield  {journal} {\bibinfo  {journal} {PNAS}\
  }\textbf {\bibinfo {volume} {112}},\ \bibinfo {pages} {4558} (\bibinfo {year}
  {2015})}\BibitemShut {NoStop}%
\bibitem [{\citenamefont {Goh}\ \emph {et~al.}(2013)\citenamefont {Goh},
  \citenamefont {Amazon},\ and\ \citenamefont {Feigenson}}]{Goh2013}%
  \BibitemOpen
  \bibfield  {author} {\bibinfo {author} {\bibfnamefont {S.~L.}\ \bibnamefont
  {Goh}}, \bibinfo {author} {\bibfnamefont {J.~J.}\ \bibnamefont {Amazon}}, \
  and\ \bibinfo {author} {\bibfnamefont {G.~W.}\ \bibnamefont {Feigenson}},\
  }\href {\doibase 10.1016/j.bpj.2013.01.003} {\bibfield  {journal} {\bibinfo
  {journal} {Biophysical Journal}\ }\textbf {\bibinfo {volume} {104}},\
  \bibinfo {pages} {853} (\bibinfo {year} {2013})}\BibitemShut {NoStop}%
\bibitem [{\citenamefont {Ackerman}\ and\ \citenamefont
  {Feigenson}(2015)}]{Ackerman2015}%
  \BibitemOpen
  \bibfield  {author} {\bibinfo {author} {\bibfnamefont {D.~G.}\ \bibnamefont
  {Ackerman}}\ and\ \bibinfo {author} {\bibfnamefont {G.~W.}\ \bibnamefont
  {Feigenson}},\ }\href {\doibase 10.1021/jp511083z} {\bibfield  {journal}
  {\bibinfo  {journal} {The Journal of Physical Chemistry B}\ }\textbf
  {\bibinfo {volume} {119}},\ \bibinfo {pages} {4240} (\bibinfo {year}
  {2015})}\BibitemShut {NoStop}%
\bibitem [{\citenamefont {Honerkamp-Smith}\ \emph {et~al.}(2009)\citenamefont
  {Honerkamp-Smith}, \citenamefont {Veatch},\ and\ \citenamefont
  {Keller}}]{Honerkamp-Smith2009}%
  \BibitemOpen
  \bibfield  {author} {\bibinfo {author} {\bibfnamefont {A.~R.}\ \bibnamefont
  {Honerkamp-Smith}}, \bibinfo {author} {\bibfnamefont {S.~L.}\ \bibnamefont
  {Veatch}}, \ and\ \bibinfo {author} {\bibfnamefont {S.~L.}\ \bibnamefont
  {Keller}},\ }\href {\doibase 10.1016/j.bbamem.2008.09.010} {\bibfield
  {journal} {\bibinfo  {journal} {Biochimica et Biophysica Acta -
  Biomembranes}\ }\textbf {\bibinfo {volume} {1788}},\ \bibinfo {pages} {53}
  (\bibinfo {year} {2009})}\BibitemShut {NoStop}%
\bibitem [{\citenamefont {Shlomovitz}\ and\ \citenamefont
  {Schick}(2013)}]{Shlomovitz2013}%
  \BibitemOpen
  \bibfield  {author} {\bibinfo {author} {\bibfnamefont {R.}~\bibnamefont
  {Shlomovitz}}\ and\ \bibinfo {author} {\bibfnamefont {M.}~\bibnamefont
  {Schick}},\ }\href {\doibase 10.1016/j.bpj.2013.06.053} {\bibfield  {journal}
  {\bibinfo  {journal} {Biophysical Journal}\ }\textbf {\bibinfo {volume}
  {105}},\ \bibinfo {pages} {1406} (\bibinfo {year} {2013})}\BibitemShut
  {NoStop}%
\bibitem [{\citenamefont {Heberle}\ \emph
  {et~al.}(2013{\natexlab{b}})\citenamefont {Heberle}, \citenamefont
  {Petruzielo}, \citenamefont {Pan}, \citenamefont {Drazba}, \citenamefont
  {Ku{\v{c}}erka}, \citenamefont {Standaert}, \citenamefont {Feigenson},\ and\
  \citenamefont {Katsaras}}]{Heberle2013}%
  \BibitemOpen
  \bibfield  {author} {\bibinfo {author} {\bibfnamefont {F.~A.}\ \bibnamefont
  {Heberle}}, \bibinfo {author} {\bibfnamefont {R.~S.}\ \bibnamefont
  {Petruzielo}}, \bibinfo {author} {\bibfnamefont {J.}~\bibnamefont {Pan}},
  \bibinfo {author} {\bibfnamefont {P.}~\bibnamefont {Drazba}}, \bibinfo
  {author} {\bibfnamefont {N.}~\bibnamefont {Ku{\v{c}}erka}}, \bibinfo {author}
  {\bibfnamefont {R.~F.}\ \bibnamefont {Standaert}}, \bibinfo {author}
  {\bibfnamefont {G.~W.}\ \bibnamefont {Feigenson}}, \ and\ \bibinfo {author}
  {\bibfnamefont {J.}~\bibnamefont {Katsaras}},\ }\href {\doibase
  10.1021/ja3113615} {\bibfield  {journal} {\bibinfo  {journal} {Journal of the
  American Chemical Society}\ }\textbf {\bibinfo {volume} {135}},\ \bibinfo
  {pages} {6853} (\bibinfo {year} {2013}{\natexlab{b}})}\BibitemShut {NoStop}%
\bibitem [{\citenamefont {Niemela}\ \emph {et~al.}(2007)\citenamefont
  {Niemela}, \citenamefont {Ollila}, \citenamefont {Hyvonen}, \citenamefont
  {Karttunen},\ and\ \citenamefont {Vattulainen}}]{Niemela2007}%
  \BibitemOpen
  \bibfield  {author} {\bibinfo {author} {\bibfnamefont {P.~S.}\ \bibnamefont
  {Niemela}}, \bibinfo {author} {\bibfnamefont {S.}~\bibnamefont {Ollila}},
  \bibinfo {author} {\bibfnamefont {M.~T.}\ \bibnamefont {Hyvonen}}, \bibinfo
  {author} {\bibfnamefont {M.}~\bibnamefont {Karttunen}}, \ and\ \bibinfo
  {author} {\bibfnamefont {I.}~\bibnamefont {Vattulainen}},\ }\href {\doibase
  10.1371/journal. pcbi.0030034} {\bibfield  {journal} {\bibinfo  {journal}
  {PLoS Computational Biology}\ }\textbf {\bibinfo {volume} {3}},\ \bibinfo
  {pages} {0304} (\bibinfo {year} {2007})}\BibitemShut {NoStop}%
\bibitem [{\citenamefont {R{\'{o}}g}\ \emph {et~al.}(2009)\citenamefont
  {R{\'{o}}g}, \citenamefont {Pasenkiewicz-Gierula}, \citenamefont
  {Vattulainen},\ and\ \citenamefont {Karttunen}}]{Rog2009}%
  \BibitemOpen
  \bibfield  {author} {\bibinfo {author} {\bibfnamefont {T.}~\bibnamefont
  {R{\'{o}}g}}, \bibinfo {author} {\bibfnamefont {M.}~\bibnamefont
  {Pasenkiewicz-Gierula}}, \bibinfo {author} {\bibfnamefont {I.}~\bibnamefont
  {Vattulainen}}, \ and\ \bibinfo {author} {\bibfnamefont {M.}~\bibnamefont
  {Karttunen}},\ }\href {\doibase 10.1016/j.bbamem.2008.08.022} {\bibfield
  {journal} {\bibinfo  {journal} {Biochimica et Biophysica Acta -
  Biomembranes}\ }\textbf {\bibinfo {volume} {1788}},\ \bibinfo {pages} {97}
  (\bibinfo {year} {2009})}\BibitemShut {NoStop}%
\bibitem [{\citenamefont {Perlmutter}\ and\ \citenamefont
  {Sachs}(2009)}]{Perlmutter2009}%
  \BibitemOpen
  \bibfield  {author} {\bibinfo {author} {\bibfnamefont {J.~D.}\ \bibnamefont
  {Perlmutter}}\ and\ \bibinfo {author} {\bibfnamefont {J.~N.}\ \bibnamefont
  {Sachs}},\ }\href {\doibase 10.1021/ja9079258} {\bibfield  {journal}
  {\bibinfo  {journal} {Journal of the American Chemical Society}\ }\textbf
  {\bibinfo {volume} {131}},\ \bibinfo {pages} {16362} (\bibinfo {year}
  {2009})}\BibitemShut {NoStop}%
\bibitem [{\citenamefont {Pandit}\ \emph {et~al.}(2004)\citenamefont {Pandit},
  \citenamefont {Vasudevan}, \citenamefont {Chiu}, \citenamefont {{Jay Mashl}},
  \citenamefont {Jakobsson},\ and\ \citenamefont {Scott}}]{Pandit2004}%
  \BibitemOpen
  \bibfield  {author} {\bibinfo {author} {\bibfnamefont {S.~A.}\ \bibnamefont
  {Pandit}}, \bibinfo {author} {\bibfnamefont {S.}~\bibnamefont {Vasudevan}},
  \bibinfo {author} {\bibfnamefont {S.}~\bibnamefont {Chiu}}, \bibinfo {author}
  {\bibfnamefont {R.}~\bibnamefont {{Jay Mashl}}}, \bibinfo {author}
  {\bibfnamefont {E.}~\bibnamefont {Jakobsson}}, \ and\ \bibinfo {author}
  {\bibfnamefont {H.}~\bibnamefont {Scott}},\ }\href {\doibase
  10.1529/biophysj.104.041939} {\bibfield  {journal} {\bibinfo  {journal}
  {Biophysical Journal}\ }\textbf {\bibinfo {volume} {87}},\ \bibinfo {pages}
  {1092} (\bibinfo {year} {2004})}\BibitemShut {NoStop}%
\bibitem [{\citenamefont {Sodt}\ \emph {et~al.}(2015)\citenamefont {Sodt},
  \citenamefont {Pastor},\ and\ \citenamefont {Lyman}}]{Sodt2015}%
  \BibitemOpen
  \bibfield  {author} {\bibinfo {author} {\bibfnamefont {A.~J.}\ \bibnamefont
  {Sodt}}, \bibinfo {author} {\bibfnamefont {R.~W.}\ \bibnamefont {Pastor}}, \
  and\ \bibinfo {author} {\bibfnamefont {E.}~\bibnamefont {Lyman}},\ }\href
  {\doibase 10.1016/j.bpj.2015.07.036} {\bibfield  {journal} {\bibinfo
  {journal} {Biophysical Journal}\ }\textbf {\bibinfo {volume} {109}},\
  \bibinfo {pages} {948} (\bibinfo {year} {2015})}\BibitemShut {NoStop}%
\bibitem [{\citenamefont {Marrink}\ \emph {et~al.}(2004)\citenamefont
  {Marrink}, \citenamefont {Vries},\ and\ \citenamefont {Mark}}]{Marrink2004}%
  \BibitemOpen
  \bibfield  {author} {\bibinfo {author} {\bibfnamefont {S.~J.}\ \bibnamefont
  {Marrink}}, \bibinfo {author} {\bibfnamefont {A.~D.}\ \bibnamefont {Vries}},
  \ and\ \bibinfo {author} {\bibfnamefont {A.}~\bibnamefont {Mark}},\ }\href
  {\doibase 10.1021/jp036508g} {\bibfield  {journal} {\bibinfo  {journal} {The
  Journal of Physical Chemistry B}\ }\textbf {\bibinfo {volume} {108}},\
  \bibinfo {pages} {750} (\bibinfo {year} {2004})}\BibitemShut {NoStop}%
\bibitem [{\citenamefont {Marrink}\ \emph {et~al.}(2007)\citenamefont
  {Marrink}, \citenamefont {Risselada}, \citenamefont {Yefimov}, \citenamefont
  {Tieleman},\ and\ \citenamefont {{De Vries}}}]{Marrink2007}%
  \BibitemOpen
  \bibfield  {author} {\bibinfo {author} {\bibfnamefont {S.~J.}\ \bibnamefont
  {Marrink}}, \bibinfo {author} {\bibfnamefont {H.~J.}\ \bibnamefont
  {Risselada}}, \bibinfo {author} {\bibfnamefont {S.}~\bibnamefont {Yefimov}},
  \bibinfo {author} {\bibfnamefont {D.~P.}\ \bibnamefont {Tieleman}}, \ and\
  \bibinfo {author} {\bibfnamefont {A.~H.}\ \bibnamefont {{De Vries}}},\ }\href
  {\doibase 10.1021/jp071097f} {\bibfield  {journal} {\bibinfo  {journal}
  {Journal of Physical Chemistry B}\ }\textbf {\bibinfo {volume} {111}},\
  \bibinfo {pages} {7812} (\bibinfo {year} {2007})}\BibitemShut {NoStop}%
\bibitem [{\citenamefont {Risselada}\ and\ \citenamefont
  {Marrink}(2008)}]{Risselada2008}%
  \BibitemOpen
  \bibfield  {author} {\bibinfo {author} {\bibfnamefont {H.~J.}\ \bibnamefont
  {Risselada}}\ and\ \bibinfo {author} {\bibfnamefont {S.~J.}\ \bibnamefont
  {Marrink}},\ }\href {\doibase 10.1073/pnas.0807527105} {\bibfield  {journal}
  {\bibinfo  {journal} {Proceedings of the National Academy of Sciences of the
  United States of America}\ }\textbf {\bibinfo {volume} {105}},\ \bibinfo
  {pages} {17367} (\bibinfo {year} {2008})}\BibitemShut {NoStop}%
\bibitem [{\citenamefont {Rosetti}\ and\ \citenamefont
  {Pastorino}(2012)}]{Rosetti2012}%
  \BibitemOpen
  \bibfield  {author} {\bibinfo {author} {\bibfnamefont {C.}~\bibnamefont
  {Rosetti}}\ and\ \bibinfo {author} {\bibfnamefont {C.}~\bibnamefont
  {Pastorino}},\ }\href {\doibase 10.1021/jp212406u} {\bibfield  {journal}
  {\bibinfo  {journal} {The Journal of Physical Chemistry B}\ }\textbf
  {\bibinfo {volume} {116}},\ \bibinfo {pages} {3525} (\bibinfo {year}
  {2012})}\BibitemShut {NoStop}%
\bibitem [{\citenamefont {Baoukina}\ \emph {et~al.}(2013)\citenamefont
  {Baoukina}, \citenamefont {Mendez-Villuendas}, \citenamefont {Bennett},\ and\
  \citenamefont {Tieleman}}]{Baoukina2013}%
  \BibitemOpen
  \bibfield  {author} {\bibinfo {author} {\bibfnamefont {S.}~\bibnamefont
  {Baoukina}}, \bibinfo {author} {\bibfnamefont {E.}~\bibnamefont
  {Mendez-Villuendas}}, \bibinfo {author} {\bibfnamefont {W.~F.~D.}\
  \bibnamefont {Bennett}}, \ and\ \bibinfo {author} {\bibfnamefont {D.~P.}\
  \bibnamefont {Tieleman}},\ }\href {\doibase 10.1039/c2fd20117h} {\bibfield
  {journal} {\bibinfo  {journal} {Faraday Discussions}\ }\textbf {\bibinfo
  {volume} {161}},\ \bibinfo {pages} {63} (\bibinfo {year} {2013})}\BibitemShut
  {NoStop}%
\bibitem [{\citenamefont {Davis}\ \emph {et~al.}(2013)\citenamefont {Davis},
  \citenamefont {{Sunil Kumar}}, \citenamefont {Sperotto},\ and\ \citenamefont
  {Laradji}}]{Davis2013}%
  \BibitemOpen
  \bibfield  {author} {\bibinfo {author} {\bibfnamefont {R.~S.}\ \bibnamefont
  {Davis}}, \bibinfo {author} {\bibfnamefont {P.~B.}\ \bibnamefont {{Sunil
  Kumar}}}, \bibinfo {author} {\bibfnamefont {M.~M.}\ \bibnamefont {Sperotto}},
  \ and\ \bibinfo {author} {\bibfnamefont {M.}~\bibnamefont {Laradji}},\ }\href
  {\doibase 10.1021/jp4000686} {\bibfield  {journal} {\bibinfo  {journal}
  {Journal of Physical Chemistry B}\ }\textbf {\bibinfo {volume} {117}},\
  \bibinfo {pages} {4072} (\bibinfo {year} {2013})}\BibitemShut {NoStop}%
\bibitem [{\citenamefont {Berendsen}\ \emph {et~al.}(1995)\citenamefont
  {Berendsen}, \citenamefont {van~der Spoel},\ and\ \citenamefont {van
  Drunen}}]{Berendsen1995}%
  \BibitemOpen
  \bibfield  {author} {\bibinfo {author} {\bibfnamefont {H.~J.~C.}\
  \bibnamefont {Berendsen}}, \bibinfo {author} {\bibfnamefont {D.}~\bibnamefont
  {van~der Spoel}}, \ and\ \bibinfo {author} {\bibfnamefont {R.}~\bibnamefont
  {van Drunen}},\ }\href {\doibase 10.1016/0010-4655(95)00042-E} {\bibfield
  {journal} {\bibinfo  {journal} {Computer Physics Communications}\ }\textbf
  {\bibinfo {volume} {91}},\ \bibinfo {pages} {43} (\bibinfo {year}
  {1995})}\BibitemShut {NoStop}%
\bibitem [{\citenamefont {Pronk}\ \emph {et~al.}(2013)\citenamefont {Pronk},
  \citenamefont {P{\'{a}}ll}, \citenamefont {Schulz}, \citenamefont {Larsson},
  \citenamefont {Bjelkmar}, \citenamefont {Apostolov}, \citenamefont {Shirts},
  \citenamefont {Smith}, \citenamefont {Kasson}, \citenamefont {{Van Der
  Spoel}}, \citenamefont {Hess},\ and\ \citenamefont {Lindahl}}]{Pronk2013}%
  \BibitemOpen
  \bibfield  {author} {\bibinfo {author} {\bibfnamefont {S.}~\bibnamefont
  {Pronk}}, \bibinfo {author} {\bibfnamefont {S.}~\bibnamefont {P{\'{a}}ll}},
  \bibinfo {author} {\bibfnamefont {R.}~\bibnamefont {Schulz}}, \bibinfo
  {author} {\bibfnamefont {P.}~\bibnamefont {Larsson}}, \bibinfo {author}
  {\bibfnamefont {P.}~\bibnamefont {Bjelkmar}}, \bibinfo {author}
  {\bibfnamefont {R.}~\bibnamefont {Apostolov}}, \bibinfo {author}
  {\bibfnamefont {M.~R.}\ \bibnamefont {Shirts}}, \bibinfo {author}
  {\bibfnamefont {J.~C.}\ \bibnamefont {Smith}}, \bibinfo {author}
  {\bibfnamefont {P.~M.}\ \bibnamefont {Kasson}}, \bibinfo {author}
  {\bibfnamefont {D.}~\bibnamefont {{Van Der Spoel}}}, \bibinfo {author}
  {\bibfnamefont {B.}~\bibnamefont {Hess}}, \ and\ \bibinfo {author}
  {\bibfnamefont {E.}~\bibnamefont {Lindahl}},\ }\href {\doibase
  10.1093/bioinformatics/btt055} {\bibfield  {journal} {\bibinfo  {journal}
  {Bioinformatics}\ }\textbf {\bibinfo {volume} {29}},\ \bibinfo {pages} {845}
  (\bibinfo {year} {2013})}\BibitemShut {NoStop}%
\bibitem [{\citenamefont {van~der Spoel}\ \emph {et~al.}(2010)\citenamefont
  {van~der Spoel}, \citenamefont {Lindahl}, \citenamefont {Hess}, \citenamefont
  {van Buuren}, \citenamefont {Apol}, \citenamefont {Meulenhoff}, \citenamefont
  {Tieleman}, \citenamefont {Sijbers}, \citenamefont {Feenstra}, \citenamefont
  {van Drunen},\ and\ \citenamefont {Berendsen}}]{gromacs456}%
  \BibitemOpen
  \bibfield  {author} {\bibinfo {author} {\bibfnamefont {D.}~\bibnamefont
  {van~der Spoel}}, \bibinfo {author} {\bibfnamefont {E.}~\bibnamefont
  {Lindahl}}, \bibinfo {author} {\bibfnamefont {B.}~\bibnamefont {Hess}},
  \bibinfo {author} {\bibfnamefont {A.~R.}\ \bibnamefont {van Buuren}},
  \bibinfo {author} {\bibfnamefont {E.}~\bibnamefont {Apol}}, \bibinfo {author}
  {\bibfnamefont {P.~J.}\ \bibnamefont {Meulenhoff}}, \bibinfo {author}
  {\bibfnamefont {D.~P.}\ \bibnamefont {Tieleman}}, \bibinfo {author}
  {\bibfnamefont {A.~L. T.~M.}\ \bibnamefont {Sijbers}}, \bibinfo {author}
  {\bibfnamefont {K.~A.}\ \bibnamefont {Feenstra}}, \bibinfo {author}
  {\bibfnamefont {R.}~\bibnamefont {van Drunen}}, \ and\ \bibinfo {author}
  {\bibfnamefont {H.~J.~C.}\ \bibnamefont {Berendsen}},\ }\href@noop {} {\emph
  {\bibinfo {title} {{GROMACS User Manual Version 4.6.5}}}}\ (\bibinfo {year}
  {2010})\BibitemShut {NoStop}%
\bibitem [{\citenamefont {Bussi}\ \emph {et~al.}(2007)\citenamefont {Bussi},
  \citenamefont {Donadio},\ and\ \citenamefont {Parrinello}}]{Bussi2007}%
  \BibitemOpen
  \bibfield  {author} {\bibinfo {author} {\bibfnamefont {G.}~\bibnamefont
  {Bussi}}, \bibinfo {author} {\bibfnamefont {D.}~\bibnamefont {Donadio}}, \
  and\ \bibinfo {author} {\bibfnamefont {M.}~\bibnamefont {Parrinello}},\
  }\href {\doibase 10.1063/1.2408420} {\bibfield  {journal} {\bibinfo
  {journal} {The Journal of Chemical Physics}\ }\textbf {\bibinfo {volume}
  {126}},\ \bibinfo {pages} {014101} (\bibinfo {year} {2007})}\BibitemShut
  {NoStop}%
\bibitem [{\citenamefont {Wassenaar}\ \emph {et~al.}(2015)\citenamefont
  {Wassenaar}, \citenamefont {Ing{\'{o}}lfsson}, \citenamefont {Bockmann},
  \citenamefont {Tieleman},\ and\ \citenamefont {Marrink}}]{Wassenaar2015}%
  \BibitemOpen
  \bibfield  {author} {\bibinfo {author} {\bibfnamefont {T.~A.}\ \bibnamefont
  {Wassenaar}}, \bibinfo {author} {\bibfnamefont {H.~I.}\ \bibnamefont
  {Ing{\'{o}}lfsson}}, \bibinfo {author} {\bibfnamefont {R.~A.}\ \bibnamefont
  {Bockmann}}, \bibinfo {author} {\bibfnamefont {D.~P.}\ \bibnamefont
  {Tieleman}}, \ and\ \bibinfo {author} {\bibfnamefont {S.~J.}\ \bibnamefont
  {Marrink}},\ }\href {\doibase 10.1021/acs.jctc.5b00209} {\bibfield  {journal}
  {\bibinfo  {journal} {Journal of Chemical Theory and Computation}\ }\textbf
  {\bibinfo {volume} {11}},\ \bibinfo {pages} {2144} (\bibinfo {year}
  {2015})}\BibitemShut {NoStop}%
\bibitem [{\citenamefont {Gowers}\ \emph {et~al.}(2016)\citenamefont {Gowers},
  \citenamefont {Linke}, \citenamefont {Barnoud}, \citenamefont {Reddy},
  \citenamefont {Melo}, \citenamefont {Seyler}, \citenamefont {Doma{\'{n}}ski},
  \citenamefont {Dotson}, \citenamefont {Buchoux}, \citenamefont {Kenney},\
  and\ \citenamefont {Beckstein}}]{Gowers2016}%
  \BibitemOpen
  \bibfield  {author} {\bibinfo {author} {\bibfnamefont {R.~J.}\ \bibnamefont
  {Gowers}}, \bibinfo {author} {\bibfnamefont {M.}~\bibnamefont {Linke}},
  \bibinfo {author} {\bibfnamefont {J.}~\bibnamefont {Barnoud}}, \bibinfo
  {author} {\bibfnamefont {T.~J.~E.}\ \bibnamefont {Reddy}}, \bibinfo {author}
  {\bibfnamefont {M.~N.}\ \bibnamefont {Melo}}, \bibinfo {author}
  {\bibfnamefont {S.~L.}\ \bibnamefont {Seyler}}, \bibinfo {author}
  {\bibfnamefont {J.}~\bibnamefont {Doma{\'{n}}ski}}, \bibinfo {author}
  {\bibfnamefont {D.~L.}\ \bibnamefont {Dotson}}, \bibinfo {author}
  {\bibfnamefont {S.}~\bibnamefont {Buchoux}}, \bibinfo {author} {\bibfnamefont
  {I.~M.}\ \bibnamefont {Kenney}}, \ and\ \bibinfo {author} {\bibfnamefont
  {O.}~\bibnamefont {Beckstein}},\ }\href
  {http://conference.scipy.org/proceedings/scipy2016/pdfs/oliver{\_}beckstein.pdf}
  {\bibfield  {journal} {\bibinfo  {journal} {Proceedings of the 15th Python in
  Science Conference}\ ,\ \bibinfo {pages} {98}} (\bibinfo {year}
  {2016})}\BibitemShut {NoStop}%
\bibitem [{\citenamefont {Hansen}\ and\ \citenamefont
  {McDonald}(2013)}]{HansenMcDonald2013}%
  \BibitemOpen
  \bibfield  {author} {\bibinfo {author} {\bibfnamefont {J.-P.}\ \bibnamefont
  {Hansen}}\ and\ \bibinfo {author} {\bibfnamefont {I.~R.}\ \bibnamefont
  {McDonald}},\ }\href@noop {} {\emph {\bibinfo {title} {Theory of Simple
  Liquids}}},\ \bibinfo {edition} {4th}\ ed.\ (\bibinfo  {publisher} {Academic
  Press},\ \bibinfo {year} {2013})\BibitemShut {NoStop}%
\bibitem [{\citenamefont {Das}\ \emph {et~al.}(2003)\citenamefont {Das},
  \citenamefont {Horbach},\ and\ \citenamefont {Binder}}]{Das2003}%
  \BibitemOpen
  \bibfield  {author} {\bibinfo {author} {\bibfnamefont {S.~K.}\ \bibnamefont
  {Das}}, \bibinfo {author} {\bibfnamefont {J.}~\bibnamefont {Horbach}}, \ and\
  \bibinfo {author} {\bibfnamefont {K.}~\bibnamefont {Binder}},\ }\href
  {\doibase 10.1063/1.1580106} {\bibfield  {journal} {\bibinfo  {journal} {The
  Journal of Chemical Physics}\ }\textbf {\bibinfo {volume} {119}},\ \bibinfo
  {pages} {1547} (\bibinfo {year} {2003})}\BibitemShut {NoStop}%
\bibitem [{\citenamefont {Pedregosa}\ \emph {et~al.}(2011)\citenamefont
  {Pedregosa}, \citenamefont {Weiss},\ and\ \citenamefont
  {Brucher}}]{pedregosa2011}%
  \BibitemOpen
  \bibfield  {author} {\bibinfo {author} {\bibfnamefont {F.}~\bibnamefont
  {Pedregosa}}, \bibinfo {author} {\bibfnamefont {R.}~\bibnamefont {Weiss}}, \
  and\ \bibinfo {author} {\bibfnamefont {M.}~\bibnamefont {Brucher}},\
  }\href@noop {} {\ \textbf {\bibinfo {volume} {12}},\ \bibinfo {pages} {2825}
  (\bibinfo {year} {2011})}\BibitemShut {NoStop}%
\bibitem [{\citenamefont {Kass}\ and\ \citenamefont
  {Raftery}(1995)}]{Kass1995}%
  \BibitemOpen
  \bibfield  {author} {\bibinfo {author} {\bibfnamefont {R.~E.}\ \bibnamefont
  {Kass}}\ and\ \bibinfo {author} {\bibfnamefont {A.~E.}\ \bibnamefont
  {Raftery}},\ }\href@noop {} {\bibfield  {journal} {\bibinfo  {journal}
  {Journal of the American Statistical Association}\ }\textbf {\bibinfo
  {volume} {90}},\ \bibinfo {pages} {773} (\bibinfo {year} {1995})}\BibitemShut
  {NoStop}%
\bibitem [{\citenamefont {Armstrong}\ \emph {et~al.}(2013)\citenamefont
  {Armstrong}, \citenamefont {Marquardt}, \citenamefont {Dies}, \citenamefont
  {Kucerka}, \citenamefont {Yamani}, \citenamefont {Harroun}, \citenamefont
  {Katsaras}, \citenamefont {Shi},\ and\ \citenamefont
  {Rheinst{\"a}dter}}]{Armstrong2013}%
  \BibitemOpen
  \bibfield  {author} {\bibinfo {author} {\bibfnamefont {C.~L.}\ \bibnamefont
  {Armstrong}}, \bibinfo {author} {\bibfnamefont {D.}~\bibnamefont
  {Marquardt}}, \bibinfo {author} {\bibfnamefont {H.}~\bibnamefont {Dies}},
  \bibinfo {author} {\bibfnamefont {N.}~\bibnamefont {Kucerka}}, \bibinfo
  {author} {\bibfnamefont {Z.}~\bibnamefont {Yamani}}, \bibinfo {author}
  {\bibfnamefont {T.~A.}\ \bibnamefont {Harroun}}, \bibinfo {author}
  {\bibfnamefont {J.}~\bibnamefont {Katsaras}}, \bibinfo {author}
  {\bibfnamefont {A.-C.}\ \bibnamefont {Shi}}, \ and\ \bibinfo {author}
  {\bibfnamefont {M.~C.}\ \bibnamefont {Rheinst{\"a}dter}},\ }\href {\doibase
  10.1371/journal.pone.0066162} {\bibfield  {journal} {\bibinfo  {journal}
  {PLoS One}\ }\textbf {\bibinfo {volume} {8}},\ \bibinfo {pages} {e66162}
  (\bibinfo {year} {2013})}\BibitemShut {NoStop}%
\bibitem [{\citenamefont {Schick}(2012)}]{Schick2012}%
  \BibitemOpen
  \bibfield  {author} {\bibinfo {author} {\bibfnamefont {M.}~\bibnamefont
  {Schick}},\ }\href {\doibase 10.1103/PhysRevE.85.031902} {\bibfield
  {journal} {\bibinfo  {journal} {Phys. Rev. E}\ }\textbf {\bibinfo {volume}
  {85}},\ \bibinfo {pages} {031902} (\bibinfo {year} {2012})}\BibitemShut
  {NoStop}%
\bibitem [{\citenamefont {Shlomovitz}\ \emph {et~al.}(2014)\citenamefont
  {Shlomovitz}, \citenamefont {Maibaum},\ and\ \citenamefont
  {Schick}}]{Shlomovitz2014}%
  \BibitemOpen
  \bibfield  {author} {\bibinfo {author} {\bibfnamefont {R.}~\bibnamefont
  {Shlomovitz}}, \bibinfo {author} {\bibfnamefont {L.}~\bibnamefont {Maibaum}},
  \ and\ \bibinfo {author} {\bibfnamefont {M.}~\bibnamefont {Schick}},\ }\href
  {\doibase 10.1016/j.bpj.2014.03.017} {\bibfield  {journal} {\bibinfo
  {journal} {Biophys. J.}\ }\textbf {\bibinfo {volume} {106}},\ \bibinfo
  {pages} {1979} (\bibinfo {year} {2014})}\BibitemShut {NoStop}%
\bibitem [{\citenamefont {Pantelopulos}\ \emph {et~al.}(2017)\citenamefont
  {Pantelopulos}, \citenamefont {Nagai}, \citenamefont {Bandara}, \citenamefont
  {Panahi},\ and\ \citenamefont {Straub}}]{Pantelopulos17}%
  \BibitemOpen
  \bibfield  {author} {\bibinfo {author} {\bibfnamefont {G.~A.}\ \bibnamefont
  {Pantelopulos}}, \bibinfo {author} {\bibfnamefont {T.}~\bibnamefont {Nagai}},
  \bibinfo {author} {\bibfnamefont {A.}~\bibnamefont {Bandara}}, \bibinfo
  {author} {\bibfnamefont {A.}~\bibnamefont {Panahi}}, \ and\ \bibinfo {author}
  {\bibfnamefont {J.~E.}\ \bibnamefont {Straub}},\ }\href {\doibase
  10.1063/1.4999709} {\bibfield  {journal} {\bibinfo  {journal} {J. Chem.
  Phys.}\ }\textbf {\bibinfo {volume} {147}},\ \bibinfo {pages} {095101}
  (\bibinfo {year} {2017})}\BibitemShut {NoStop}%
\bibitem [{\citenamefont {Hakobyan}\ and\ \citenamefont
  {Heuer}(2013)}]{Hakobyan2013}%
  \BibitemOpen
  \bibfield  {author} {\bibinfo {author} {\bibfnamefont {D.}~\bibnamefont
  {Hakobyan}}\ and\ \bibinfo {author} {\bibfnamefont {A.}~\bibnamefont
  {Heuer}},\ }\href {\doibase 10.1021/jp312245y} {\bibfield  {journal}
  {\bibinfo  {journal} {The Journal of Physical Chemistry B}\ }\textbf
  {\bibinfo {volume} {117}},\ \bibinfo {pages} {3841} (\bibinfo {year}
  {2013})}\BibitemShut {NoStop}%
\bibitem [{\citenamefont {Bruce}(1981)}]{Bruce81}%
  \BibitemOpen
  \bibfield  {author} {\bibinfo {author} {\bibfnamefont {A.~D.}\ \bibnamefont
  {Bruce}},\ }\href {\doibase 10.1088/0022-3719/14/25/012} {\bibfield
  {journal} {\bibinfo  {journal} {J. Phys. C}\ }\textbf {\bibinfo {volume}
  {14}},\ \bibinfo {pages} {3667} (\bibinfo {year} {1981})}\BibitemShut
  {NoStop}%
\bibitem [{\citenamefont {Binder}(1981)}]{Binder81c}%
  \BibitemOpen
  \bibfield  {author} {\bibinfo {author} {\bibfnamefont {K.}~\bibnamefont
  {Binder}},\ }\href {\doibase 10.1103/PhysRevLett.47.693} {\bibfield
  {journal} {\bibinfo  {journal} {Phys. Rev. Lett.}\ }\textbf {\bibinfo
  {volume} {47}},\ \bibinfo {pages} {693} (\bibinfo {year} {1981})}\BibitemShut
  {NoStop}%
\end{thebibliography}%

\end{document}